\documentclass[10pt,journal]{IEEEJtran}
\ifCLASSOPTIONcompsoc
  \usepackage[nocompress]{cite}
\else
  \usepackage{cite}
\fi
\ifCLASSINFOpdf
\else
\fi
\usepackage{color, xcolor, float, lscape, enumerate, graphicx, url, tabularx, multirow, xspace}
\usepackage[bookmarks=false]{hyperref}

\usepackage{caption}
\usepackage{verbatim}
\hypersetup{  
	pdfpagemode=pagewidth,
	plainpages=false, 
	colorlinks,  
	urlcolor=red!50!black,   
	linkcolor=blue!50!black,    
	citecolor=blue!50!black,  
	bookmarksnumbered
}

\usepackage[utf8]{inputenc}
\usepackage{comment}
\usepackage{blindtext}
\usepackage{graphicx}
\usepackage{booktabs} 
\usepackage{amsfonts}
\usepackage{amsmath} 
\usepackage{booktabs}
\usepackage{multirow}
\usepackage{graphicx}
\usepackage{graphics}
\usepackage{subfigure}
\usepackage{balance}
\usepackage{xcolor}
\usepackage{color}
\usepackage{tikz,pgf}
\usepackage{hyperref}
\usepackage{booktabs} 
\usepackage{enumitem}
\usepackage{float}
\usepackage{xspace}
\usepackage{blindtext}
\newcommand{\etal}{{\em et al.}\xspace}
\newcommand{\eg}{{\em e.g.,}\xspace}
\newcommand{\ie}{{\em i.e.,}\xspace}
\newcommand{\etc}{{\em etc.}\xspace}
\newcommand{\BfPara}[1]{{\noindent\bf#1.}\xspace}
\usepackage{pifont}
\newcommand{\xmark}{\ding{55}}
\newcommand{\cmark}{\ding{51}}

\newcommand*\cib[1]{\tikz[baseline=(char.base)]{
                            \node[shape=circle,color=black, fill=black,text=white,draw,inner sep=0.3pt] (char) {#1};}}

\usepackage{xcolor,colortbl}

\definecolor{darkgreen}{rgb}{0.0, 0.2, 0.13}

\newcommand{\cc}[1]{\cellcolor{darkgreen!50!white!#1}}
\newcommand{\err}[1]{\cellcolor{red!60!white!#1}}

\begin{document}
\title{Sensor-based Continuous Authentication of Smartphones' Users Using Behavioral Biometrics: A Contemporary Survey}
\markboth{IEEE INTERNET OF THINGS JOURNAL}%
{Abuhamad~\etal Survey: Sensor-based Continuous Authentication of Smartphones' Users Using Behavioral Biometrics}

\author{Mohammed~Abuhamad,
        Ahmed~Abusnaina,
        DaeHun~Nyang,
        and~David~Mohaisen%
\IEEEcompsocitemizethanks{\IEEEcompsocthanksitem M. Abuhamad, A. Abusnaina, and D. Mohaisen are with the Department of Computer
Science, University of Central Florida, Orlando, FL 32816. D. Nyang is with the Department of Computer Information Engineering, Ewha Womans University, Seoul, South Korea
E-mail: \{abuhamad,ahmed.abusnaina\} @knights.ucf.edu, nyang@ewha.ac.kr, and mohaisen@cs.ucf.edu.\protect\\

}
}




\maketitle

\begin{abstract}
Mobile devices and technologies have become increasingly popular, offering comparable storage and computational capabilities to desktop computers allowing users to store and interact with sensitive and private information. 
The security and protection of such personal information are becoming more and more important since mobile devices are vulnerable to unauthorized access or theft. 
User authentication is a task of paramount importance that grants access to legitimate users at the point-of-entry and continuously through the usage session.
This task is made possible with today's smartphones' embedded sensors that enable continuous and implicit user authentication by capturing behavioral biometrics and traits.  
In this paper, we survey more than 140  recent behavioral biometric-based approaches for continuous user authentication, including motion-based methods (28 studies), gait-based methods (19 studies), keystroke dynamics-based methods (20 studies), touch gesture-based methods (29 studies),  voice-based methods (16 studies), and multimodal-based methods (34 studies). 
The survey provides an overview of the current state-of-the-art approaches for continuous user authentication using behavioral biometrics captured by smartphones' embedded sensors, including insights and open challenges for adoption, usability, and performance.
\end{abstract}


\section{Introduction}
\label{sec:introduction}
\IEEEPARstart{S}{martphones} have been witnessing a rapid increase in storage and computational resources, making them an invaluable instrument for activities on the internet and a leading platform for users' communication and interaction with data and media of different forms.
Moreover, the current edge and cloud computing services available to users have increased the reliance on mobile devices for mobility and convenience, revolutionizing the landscape of technologies and methods of conducting transactions \cite{jung2018digitalseal}.
The continuous user authentication is an implicit process of validating the legitimate user based on capturing behavioral attributes by leveraging resources and built-in sensors of the mobile device.
Users tend to develop distinctive behavioral patterns when using mobile devices, which can be used for the authentication task. 
These patterns are implicitly captured as users interact with their devices using behavioral features calculated from a stream of data, such as interaction and environmental information and sensory data.
Continuous authentication methods are also called ``{\em transparent}, {\em implicit}, {\em active}, {\em non-intrusive}, {\em non-observable}, {\em adaptive}, {\em unobtrusive}, and {\em progressive}'' techniques~\cite{AlAbdulwahidCSFR16,neal2016surveying}. 
Traditionally, continuous authentication methods operate as a support process to the conventional authentication methods, \eg using secret-based authentication or physiological biometrics, such as prompting users to re-authenticate when adversarial or unauthorized behavior is detected.

Recently, the field of continuous authentication has been gaining increasing interest, especially with the expansion of storage and computational resources and the availability of sensors that can make the implicit authentication very accurate and effective. Using sensors-based authentication methods offers convenient and efficient access control for users. This paper surveys recent and state-of-the-art methods for continuous authentication using behavioral biometrics. We aim to shed light on current state and challenges facing the adoption of such methods in today's smartphones.

\BfPara{Conventional vs. Biometric Approaches} 
To date, vendors of mobile devices have adopted both knowledge-based schemes and physiological biometrics as the primary security method for accessing the device. 
Knowledge-based approaches rely on the knowledge of the user; \ie the user must provide certain information such as numeric password, PIN, graphical sequence, or a picture gesture \cite{zhao2015picture}, to access a device \cite{nyang2014keylogging}. 
Despite their simplicity, ease of implementation, and user acceptance, such approaches suffer from several shortcomings such as the inconvenience of frequent re-entering (especially when the knowledge used are long enough to convey strong security) and
several adversarial attacks (\eg shoulder surfing and smudge attacks) \cite{nyang2018two,de2012touch,clarke2005authentication,amin2014biometric,crawford2014understanding}. 
Another issue with knowledge-based authentication is the underlying assumption of having equal security requirements for all applications \cite{furnell2008beyond}. 
For example, accessing financial records and texting are given the same level of security. Using a knowledge-based authentication on smartphones falls short on delivering application-specific security guarantees \cite{khan2014towards}, especially observing the recent emergence of adaptable biometric authentication that account for environmental factors to adapt and select the suitable sensors for authentication (\eg using fingerprint sensor when the lighting condition does not allow for face recognition) \cite{wojtowicz2016model}.
Even when using more complicated implementations of knowledge-based approaches, \eg Yu \etal{}'s \cite{yu2016usable} implementation of 3D graphical passwords which can be easier to remember and possibly providing larger password space, they still inherit the same drawbacks. 
In fact, in a study by Amin \etal \cite{amin2014biometric},
 graphical sequences (2D patterns) are shown to be as easier to predict as textual passwords since 40\% of patterns begin from the top-left node and the majority of users use five nodes out of the nine nodes. 
Another example of sophisticated knowledge-based schemes is introduced by Shin \etal \cite{shin2012design}, which includes changing the colors of six circles by touching them repeatedly up to seven times. Once all the circles' colors fit the correct combination, user authentication is granted. Even though this allows for harder security (especially when enabling a larger number of circles and colors), it still requires memorizing such complex combinations, which is the main disadvantage in knowledge-based approaches.
To overcome the need for memorizing complex combinations, Yang \etal \cite{yang2016free} proposed free-form gestures (doodling) as a user validation scheme, where users are to enter any draw with any number of fingers. The authors showed that using free-form gestures enabled a  log-in time reduction reached 22\% in comparison to textual passwords while maintaining higher usability and search space. However, the authors have not addressed other security concerns such as shoulder surfing and smudge attacks. 

Many researchers have attempted to overcome the core problems of knowledge-based authentication by coupling such methods with biometric-based methods. Using biometric information improves both the accuracy and usability of the authentication process. Such integration can be done by measuring the keystroke dynamics or gestures when connecting, changing the order, or selecting images \cite{li2014active}. 
The shortcomings of knowledge-based authentication approaches motivate for using stronger and easier authentication schemes such as biometrics.
Physiological biometrics provide unquestionable precision of user authentication with a convenient and simple approach.
For example, most current smartphones are equipped with a fingerprint recognition module as reliability and a cost-effective method for user authentication \cite{neal2016surveying,zhou2019multi,cilia2018multi}.

Physiological biometrics-based authentication techniques
show high efficiency, accuracy, and user acceptance \cite{neal2016surveying, khan2014towards, miles2006tracking}. 
However, all of these techniques necessitate the user's knowledge of the service since the user must interact with the biometric sensor and be aware of the biometric capturing process. 
Similar to knowledge-based authentication schemes, physiological biometrics, \eg face, fingerprint, periocular, and iris, can provide point-of-entry authentication and fall short of offering implicit and transparent authentication. 

\BfPara{Motivation}
It is obvious that knowledge-based and physiological biometric-based methods are successful for user validation, but they fall short on delivering continuous and transparent authentication. Moreover, physiological biometrics are mostly hardware-dependent. 
Behavioral biometrics show higher potential to meet all requirements for an efficient authentication system. 
In addition to all benefits of adopting behavioral biometrics, they are a suitable solution for  ``user abandonment'' \cite{schaffer2015expanding} protection, or when the legitimate user of the unlocked device is not present. 
These many advantages of behavioral biometrics-based authentication have shown to be influential for user adoption since a survey by Crawford and Renaud~\cite{crawford2014understanding} demonstrated that 90\% of the study's participants favored behavioral biometrics-based transparent authentication. Hence, the literature shows a remarkable interest in adopting various behavioral modalities, such as keystroke dynamics, touch gestures, motion, voice, \etc, for transparent user authentication on mobile devices. 

This paper focuses on behavioral continuous authentication and multimodal methods that may incorporate physiological biometrics to harden security and boost the performance of the authentication scheme. For readability, we list the abbreviations used in this paper in Table \ref{tab:abbreviations}.

\BfPara{Other Related Surveys}%
There are several surveys that have addressed specific modalities, \eg keystroke dynamics~\cite{teh2013survey,6496441,karnan2011biometric}, voice-based speaker identification~\cite{tirumala2017speaker}, multimodal authentication~\cite{AlAbdulwahidCSFR16}. 
Moreover, there are surveys that address traditional biometric-based, \ie \cite{clarke2005authentication,Spolaor2016}, general authentication schemes, \ie \cite{neal2016surveying,AlAbdulwahidCSFR16}, and authentication protocols and OS-related security \cite{kunda2018survey}.
This study provides a contemporary survey for sensor-based continuous authentication on smartphones, differing in scope, time, and range of surveyed works.

\begin{table}[t]
\begin{center}
\caption{List of abbreviations in alphabetical order.}
\label{tab:abbreviations}
\begin{tabular}{l|l}
\toprule
\textbf{Term}  & \textbf{Definition} \\

\midrule
    
    Ac & Accelerometer \\
    ANN & Artificial Neural Network \\
    Ca & Camera\\
    CC & Cross-correlation \\
    CI & Confidence Interval \\
    CNN & Convolutional Neural Network \\
    Co & Compass \\
    CPANN & Counter Propagation Artificial Neural Network\\
    CRM & Cyclic Rotation Metric \\ 
    DAE-SR & Deep Auto Encoder and Softmax Regression\\
    DSP & Digital Signal Processing \\
    DTW & Dynamic Time Wrapping \\
    EEH & Electromagnetic Energy Harvester \\
    EER & Equal Error Rate \\
    El & Elevation\\
    FA-NN & Fast Approximate Nearest Neighbor \\
    FAR & False Acceptance Rate \\
    FC & Fuzzy Commitment \\
    FFT & Fast Fourier Transform \\
    FLD & Fisher Linear Discriminant \\
    FPOS & Frequent Pattern Outlier Score \\
    FRR & False Rejection Rate \\
    FSR & Force Sensing Resistor \\
    GA & Genetic Algorithm \\
    GMM & Gaussian Mixed Model \\
    GPS & Global Positioning System \\
    Gr & Gravity sensor \\
    Gy & Gyroscope \\
    HMM & Hidden Markov Model \\
    HWS & Healthcare Wearable Sensors \\
    I-F & Isolation Forest \\ 
    KL & Kullback-Leibler \\
    k-NN & k-Nearest Neighbor \\
    KRR & Kernel Ridge Regression\\
    LDA & Linear Discriminant Analysis\\
    Li & Light sensor\\
    LMC & Leap Motion Controller\\
    LSTM & Long Short Term Memory \\
    Ma & Magnetometer \\
    MCF & Multi-Classifier Fusion \\
    MGGN & Multivariate Gaussian Generative Model \\
    MHD & Modified Hausdorff Distance\\
    Mi & Microphone \\
    MLP & Multilayer Perceptron \\
    MRC & Cyclic Rotation Metric \\
    Or & Orientation \\
    PCA & Principle Component Analysis\\
    PEH & Piezoelectric Energy Harvester \\
    Pr & Pressure\\
    PSO & Particle Swarm Optimization\\
    RBF & Radial Basis Function \\
    RBFN & Radial Basis Function Network \\
    RF & Random Forest\\
    SOM & Self Organizing Maps \\
    Sp & Speaker \\
    SRC & Sparse Representation Classification\\
    SVM & Support Vector Machine \\
    To & Touch\\
    VR & Virtual Reality \\

\bottomrule
\end{tabular}
\end{center} 
\end{table}

\BfPara{Contribution} This work contributes to the mobile continuous user authentication in several aspects:
\begin{itemize}[leftmargin=*]
    \item Survey more than 140 works on continuous user authentication methods, categorizing them into six behavioral and physiological biometrics groups (motion, gait, keystroke dynamics, gesture, voice, and multimodal).
    \item Present the studies of each biometric modality in a table format, comparing works by the modality, sensors, and the used authentication algorithm, in addition to the data collected, user sample size, and six evaluation metrics. Such comparison provides ease in understanding each work and how it compares to others in the field.
    \item Give insights and challenges for different biometric methods, highlighting the possible future work and existing common gaps within the literature.
\end{itemize}

\BfPara{Organization} This survey is organized as follows: we discuss the system design of continuous user authentication, including biometric modalities, user enrollment, and verification techniques, and evaluation metrics in Section~\ref{sec:systemDesign}. 
The user authentication system is categorized into six groups: motion-based authentication is discussed in Section~\ref{sec:motion}, gait-based authentication in Section~\ref{sec:gait}, followed by keystroke dynamics-based authentication in Section~\ref{sec:keystroke}. Touch gesture-based and voice-based authentication methods are described in Section~\ref{sec:gesture} and Section~\ref{sec:voice}, respectively. The multimodal-based authentication is described in Section~\ref{sec:multimodalAuthentication}. Finally, we conclude in Section~\ref{sec:conclusion}.

\section{Continuous Authentication: Design}\label{sec:systemDesign}

Numerous studies have explored various methods for continuous user authentication leveraging modern mobile technologies and embedded sensors to model users' behavior. The deployment of sensors on today's mobile devices have enabled a variety of applications, such as modeling human behavior \cite{EhatishamANAL18,NwekeTAA18}, user authentication \cite{AmanBS19, ShenLCGM18, ZhangHXCH17, Arteaga-Falconi16,BaPFKMR18}, activity and action recognition \cite{ZengNYMZWZ14, BiegelC04, EhatishamANAL18}, and healthcare monitoring \cite{DasGD16,ChoiKCPKWK17}, among others \cite{WuJ19, YuXDGY19}. In this paper, we show recent user authentication methods that use mobile sensory data to capture users' behavioral biometrics.


\begin{figure*}[t]
    \centering
\begin{minipage}[t]{0.49\textwidth}
\includegraphics[width=0.99\textwidth]{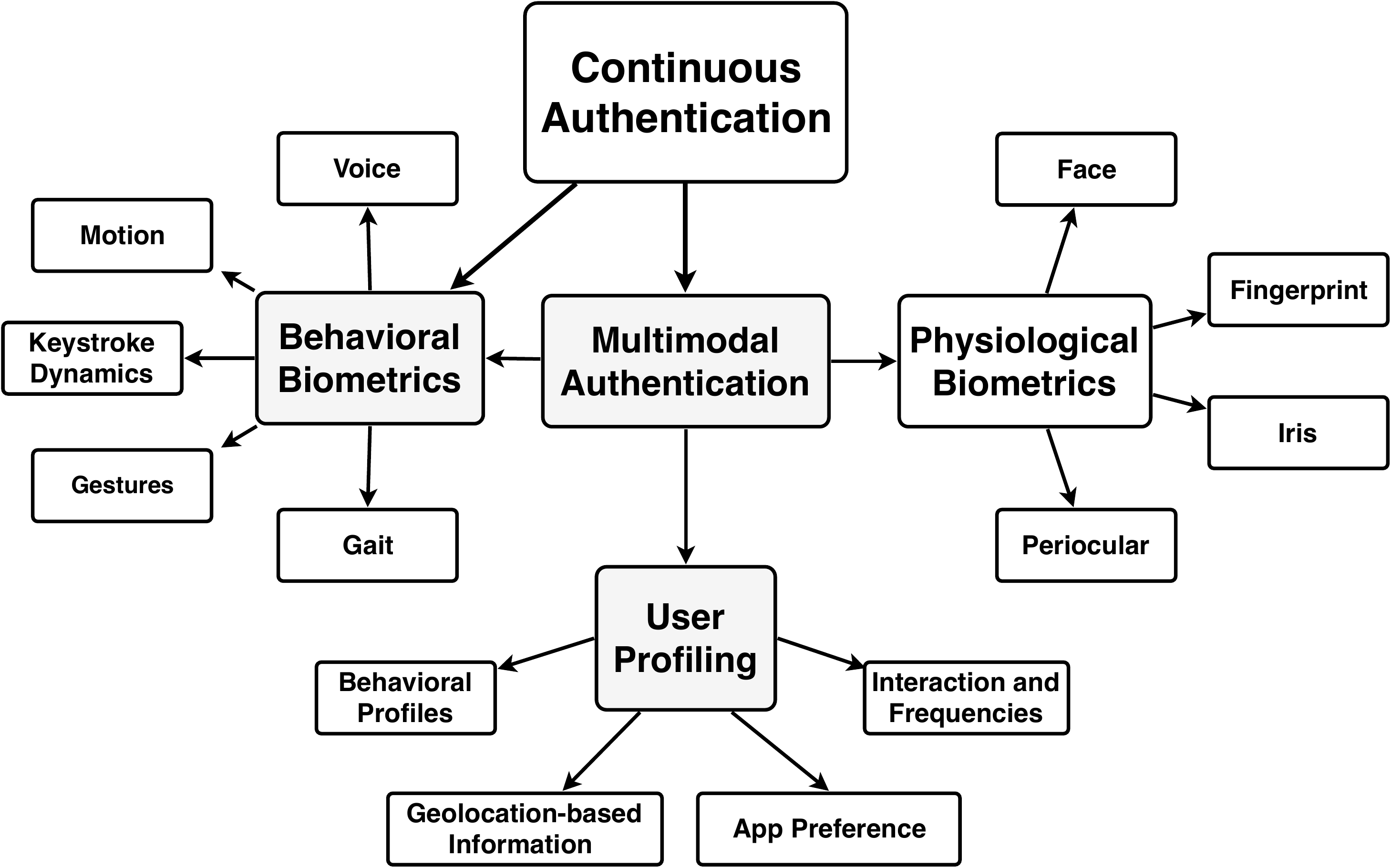}
\captionof{figure}{Biometric-based authentication modalities are categorized into physiological biometrics, behavioral biometrics, and user profiles. 
}
\label{fig:biomtricModalities}
\end{minipage}%
\hfill
\begin{minipage}[t]{0.49\textwidth}
\includegraphics[width=0.99\textwidth]{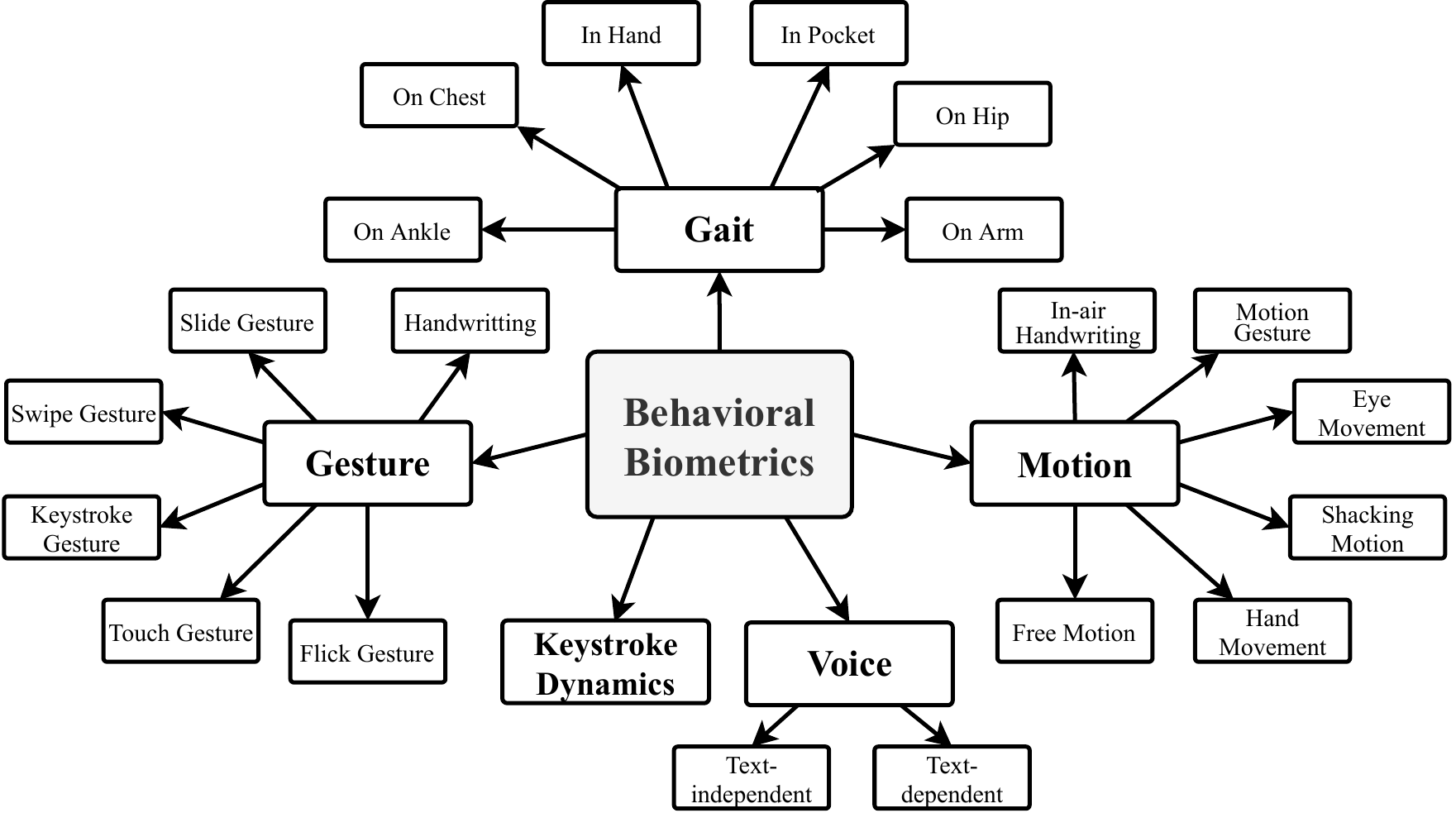}
\captionof{figure}{Behavioral biometrics are categorized into several modalities. The combination of the modalities provides a multimodal user authentication.}
\label{fig:BehavioralBiometrics}
\end{minipage}%
\end{figure*}

\subsection{Used Biometric Modalities}
Several modalities are used for biometric-based authentication, including physiological biometrics (\eg face, fingerprint, iris, \etc) and behavioral biometrics (\eg keystroke dynamics, touch gestures, voice, motions, \etc). Figure \ref{fig:biomtricModalities} shows a categorization of used modalities for user authentication tasks.
Figure \ref{fig:BehavioralBiometrics} shows the modalities and features of several behavioral biometrics that are commonly used for user authentication tasks.
All these modalities are made possible by the embedded mobile sensors, \eg camera, microphone, accelerometers, and gyroscopes, which contribute to the enrolment phase and the verification part of the authentication process.
Such sensors provide sufficient information for accurate and secure authentication, and adopting the proper utilization mechanism would play an essential role in delivering efficient and usable user authentication \cite{micallef2015sensor}. 
Using biometrics for authentication, there are enormous studies that demonstrated the benefits and security aspects of using such information to explore ``on-the-move biometry'' \cite{drosou2012activity}.

\begin{figure*}[t]
    \centering
    \includegraphics[width=0.95\textwidth]{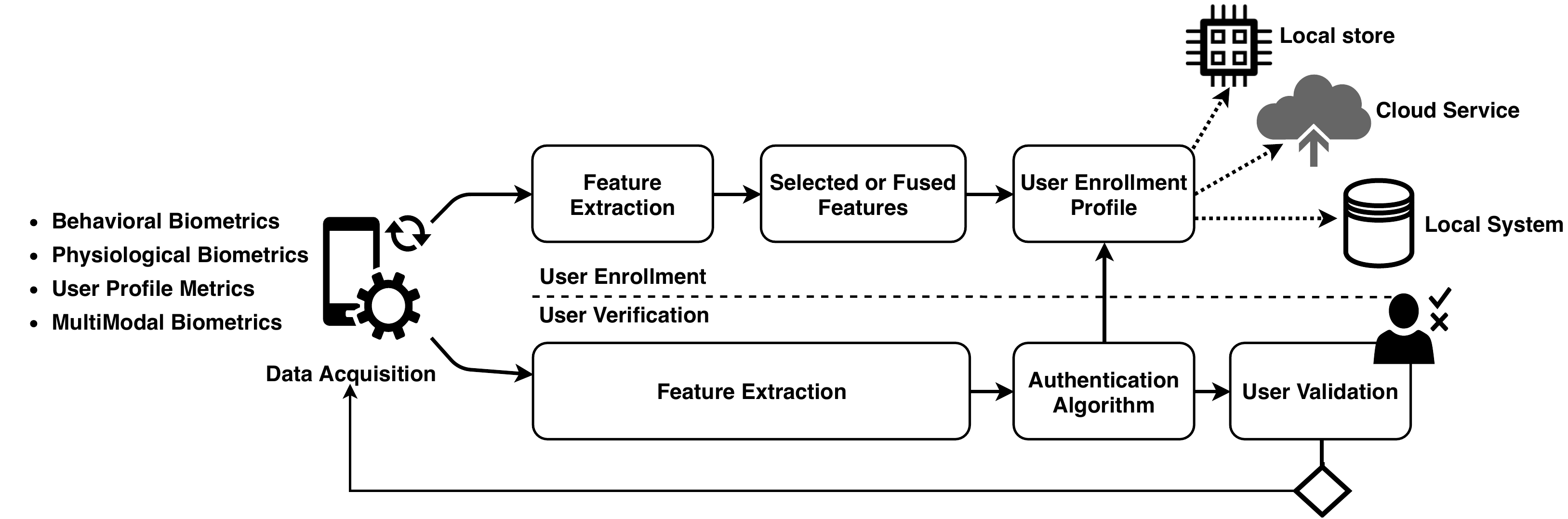}
    \caption{General framework of biometric-based authentication system. The framework includes two operations: user enrollment and user verification. Both operations require data acquisition and feature extraction. User enrollment includes modeling of the extracted data and storing, while user verification feed the extracted features to authentication algorithm to grant access for legitimate users periodically.}
    \label{fig:system}
\end{figure*}

\subsection{User Authentication}
Biometric-based user authentication leverages users' behavioral patterns for the identification or/and validation task using a pattern recognition method. 
The authentication is commonly referred to as a verification task in mobile security since the authentication method validates the legitimate user given certain biometrics. The general framework for the authentication system is illustrated in Figure \ref{fig:system}.

\BfPara{Enrolment}
There are two common approaches for user enrollment in the user authentication system. For simplicity purposes, we categorize enrollment techniques to \cib{1} template-based enrollment and \cib{2} model-based enrollment. For template-based enrollment, the user submits several samples to establish templates for future comparison. This method is popular among authentication methods using physiological biometrics, where features can be more robust to intra-class variations and more distinctive and scalable for a large population. Once users' templates are established, a similarity-based technique is used to validate users after passing a similarity threshold. Many considerations should be taken to ensure the quality of templates for supporting the performance of the system, such as the robustness and distinguishability of features across users, removing outliers, and reducing noise and redundancy. Moreover, security concerns should be addressed to ensure the security and privacy of users' templates, whether during enrollment and template registration, storing, retrieving, and processing for user authentication.
For model-based enrollment, users' biometrics are collected for training a machine learning model for user authentication, where the authentication model decides whether the input data belongs to the legitimate user. The common machine learning approaches are used to establish users' models, including data acquisition and preprocessing, feature extraction and selection, and modeling. The quality of features plays a significant role in the performance of model-based authentication. Therefore, most efficient methods include a feature evaluation and selection process to extract the most distinctive features across a large population.
Recently, model-based approaches have been gaining success for the user authentication task. However, several challenges should be tackled for efficient adoption, such as data collection size, training time, model size and robustness against possible adversarial attacks.

\BfPara{User Verification}
After the user enrollment, the system validates the legitimate user based on extracted features. The verification can be at the point-of-entry and continuously through the usage session. For continuous authentication, the user verification process occurs periodically to grant access to the legitimate user and to deny access to impostors. The frequency of verification should be carefully selected to allow sufficient biometric data acquisition and features extraction process and to manage energy consumption.
Depending on the enrolment approach, the authentication algorithm follows a similarity-based or probability-based scheme for user validation. Similarity-based techniques are used for measuring the similarity of input data in comparison to a stored template for a certain user.
Traditionally, the verification implies a match between a given data and a stored template to a certain degree. The authentication system is responsible for giving access to the legitimate user when presenting a biometric data that matches the supposed template with similarity check higher than a predefined threshold. The threshold is for accounting for environmental and processing errors that could affect the reading or calculating of the biometric data.
Mathematically, a verification process can be viewed as 
$C = True \quad \text{if} \quad f(x,y)\geq t$ and $False$ otherwise, where
 $f$ is a similarity measurement between an input  $x$ and a template $y$, and $t$ is a predefined threshold. The genuine match is shown when $C$ evaluates to $True$, while the impostor match is when $C$ is $False$.

Probability-based algorithms are used for model-based enrolment, where the authentication model signals a probability for granting access to the legitimate user based on the input data, the verification process is similar to the template-based algorithm, except for using a pre-trained model for decision making. The decision of the model $C = True \quad \text{if} \quad  g(x) \geq th$ and $False$ otherwise, where $g$ is the objective function of the probability-based algorithm and $th$ is a predefined threshold. The user verification process runs periodically for continuous user authentication, however, the frequency $f_{req}$ higher bound is limited by minimum verification time $t_o$, where $f_{req} = \frac{1}{t_o}$, and $t_o = t_d + t_p + t_c$, where $t_d$ is the time needed to acquire sufficient data for verification, $t_p$ refers to the time required for data preprocessing, and $t_c$ is classification period. While $t_c$ can be mitigated by overlapping $t_d$ and $t_p$ with $t_c$, it should be taken into account the computational power and battery consumption needed for the verification process.

\subsection{Authentication Evaluation Metrics}

Biometric-based authentication systems are evaluated by their ability to be generalized to a large population. This emphasis becomes more obvious when addressing mobile security since the authentication system should account for a very large and different population.
There are several evaluation metrics for evaluating authentication system performance. The three most common metrics are the false accept rate (FAR), the false reject rate (FRR), and the equal error rate (EER).
For the authentication task on a mobile device, a false accept indicates that false access is granted to an intruder, while a false reject indicates that the legitimate user is denied access to the device. 
FAR is represented as $\frac{\textit{Number of False Acceptance}}{\textit{Total Number of Attempts}}$ and FRR is equal to $\frac{\textit{Number of False Rejections}}{\textit{Total Number of Attempts}}$.
The EER is where the FAR is roughly similar to the FRR, and it is a very popular metric for interpreting system error. 

Additional evaluation metrics for the authentication system include true positive rate, true negative rate, false positive rate, false negative rate, accuracy, precision, recall, and F1-score. 
True positive rate and true negative rate indicate the rate of correctly validating a legitimate user and denying an impostor, respectively. 
False positive rate and the false negative rate is the rate of which the system denies access for the legitimate user and allows access for the impostor, respectively.
Accuracy is the proportion of true positives and negatives to the overall tested data, including (true positives, true negatives, false positives, and false negatives). 
Precision indicates how frequently the system correctly produces positive classifications, which is calculated as the ratio of true positives to both true and false positives. 
Recall indicates how frequently the system correctly validates positive data, which is calculated as the ratio of true positives to both true positives and false negatives.

\subsection{Behavioral Biometrics and Smartphones' Capabilities}\label{sec:physiologicalBiometrics}

Behavioral biometrics enable efficient implementation of an authentication system that operates beyond the point-of-entry access and continuously authenticate users without explicitly asking their input. Therefore, behavioral biometrics improve mobile security by providing user continuous and transparent authentication process throughout the entire routine session. 
Various techniques have been proposed for mobile user authentication using behavioral usage and features by taking advantage of the embedded sensors. Using sensory data, a background process continuously and implicitly captures user's behavior to perform an active and transparent authentication, e.g., using motion patterns \cite{DraffinZZ13, EhatishamANAL18, laghari2016biometric,ShenDXLWCW20,ShenYDXLW18,AbuhamadAMN20}, gait \cite{LopezLSS16, PozoASC12, DamaseviciusMVW16, Lu2017, NickelBB11, ZengNYMZWZ14}, touch gestures \cite{zaliva2015passive,AvivGMBS10, DraffinZZ13, WuC15, BuchouxC08,LuDWWMMS18}, electrocardiography (ECG) \cite{Arteaga-Falconi16},  
keystroke dynamics \cite{WuC15, Mondal2017, KambourakisDPP16, cilia2018multi}, voice \cite{ReynoldsQD00, LuBPKL11,gofman2016hidden}, signature \cite{ClarkeM07, Martinez-DiazFGO08, morris2006multimodal}, and profiling \cite{AlzubaidiK16, ShenLCGM18, NwekeTAA18}. 


Since today's smartphones are well-equipped with a variety of embedded sensors, such as motion sensors (\eg gravity, accelerometer, gyroscope, and magnetometer), environmental sensors (\eg light, temperature, barometer, and proximity), and position sensors (\eg GPS and compass), numerous studies have leveraged these sensors for user authentication \cite{ShenLCGM18,EhatishamANAL18,AminiNPGYK18,LeeL17}. 
A study by Crawford \etal~\cite{crawford2013framework} shows that behavioral biometrics reduce the demand for legitimate authentication by 67\% in comparison to knowledge-based methods, \ie adding a remarkable improvement in usability. 
In terms of exploiting access privilege, the authors showed that an intruder could perform more than 1,000 tasks if successfully gain access to a mobile device using a knowledge-based authentication scheme; however, the intruder can hardly achieve one task if the mobile device uses a multimodal behavioral biometrics-based method \cite{crawford2013framework}.

\begin{table*}[t]
\centering
\caption{Summary of the related work for motion-based user authentication. Each work is identified by the used modalities, utilized sensors, dataset, modeling algorithm, and their performance.}
\label{table:motion}
\scalebox{1}{\begin{tabular}{lllcccccccc}
\toprule
Study & Modalities & Sensors &  Methods & \begin{tabular}[c]{@{}c@{}}\# \\ Users\end{tabular} & EER & FAR & FRR & TPR  & Accuracy & \begin{tabular}[c]{@{}c@{}}Auth. \\ Time\end{tabular} \\
\midrule

\cite{hong2015waving}           &      Motion Gesture   &    Ac             & SVM           &  8        & \xmark & 3.67\% & \xmark &   \xmark   &\cc{93}92.83\%  &   \xmark \\
\cite{feng2013investigating}    &      Picking-up Motion   &    Ac, Gy, Ma              & SVM           &  31        & \err{6}6.13\% & \cmark & \cmark &   \xmark   &\xmark & \xmark \\
\cite{wu2015implicit} &     Motion \& keystroke &    Ac, Gy, Pr              & SVM           &  100        &\err{1} 1.25\% & \cmark & \cmark &   \xmark   & \cc{99}99.13\% &   \xmark \\
\cite{hong2016mgra} &      Motion Gesture &    Ac              & SVM           &  8        & \xmark & \xmark & \xmark &   \xmark   &  \cc{96}95.83\% & \xmark  \\
\cite{lu2018multifactor} &      In-air Handwriting &    LMC              & SVM           &  100        &\err{1} 0.6\% & \cmark & \cmark &   \xmark   &\xmark    &   \xmark \\
\cite{xia2018motionhacker} &      In-air Handwriting &    Ac, Gy              & RF           &  5        & \xmark & \xmark & \xmark &   \xmark   & \cc{33}$32.8^{\dagger}\%$ &\xmark    \\
\cite{yan2018towards} &         Shacking Motion &    Ac, Gy              & DTW-LSTM           &  150       & \xmark & 0.1\% & \xmark &   \xmark   &\cc{97}96.87\%  &   \xmark \\
\cite{fantana2015movement} &   Free-form Gesture &    Ac, Gy              & DTW           &  \xmark        & \err{3} 3\% & 0.02\% & 10\% &   \xmark   &\xmark  &   \xmark \\
\cite{casanova2010real} &      In-air Handwriting &    Ac              & $\text{DTW}^{*}$            &  34        & \err{2}2.5\% & \cmark & \cmark &   \xmark     & \xmark  &   \xmark \\
\cite{haring2018pick} &  In-air Handwriting &    Ac, Gy, Or        & MLP           &  \xmark        & \xmark & \xmark& \xmark &   84.5\%   & \xmark &     \xmark \\

\cite{laghari2016biometric} &  Pick-up Motion &    Ac              & CC           &  10        & \xmark & 1.46\%& 6.87\% &   \xmark   & \xmark &     \xmark \\

\cite{maghsoudi2016behavioral} & Motion Gesture & Ac, Gy & Na\"ive Bayes & 10 & \xmark & \xmark& \xmark& \xmark & \cc{84} 83.6\% & \xmark \\

 &  &  & k-NN & 10 & \xmark & \xmark& \xmark& \xmark & \cc{90}89.8\% & \xmark \\

& &  & MLP & 10 & \xmark & \xmark& \xmark& \xmark & \cc{93}92.7\% & \xmark \\

& &  & SVM & 10 & \xmark & \xmark& \xmark& \xmark & \cc{92}92.2\% & \xmark \\

\cite{eremin2018touch} & Hand-movement & Ac, Gy & Na\"ive Bayes & 50 & \xmark & 2\% & \xmark& \xmark & \cc{89} 89\% & \xmark \\

& & Ma, Or, Gy & SVM & 50 & \xmark & 18\% & \xmark& \xmark & \cc{74}74\% & \xmark \\

& &   & I-F & 50 & \xmark & 0\% & \xmark& \xmark & \cc{93}93\% & \xmark \\

\cite{LeeL15} &  Free motion &    Ac, Ma, Or        & SVM          &  4        & \xmark & \xmark& \xmark &   \xmark  & \cc{90}90\% &     20s \\

\cite{EhatishamANAL18} &  Free motion &    Ac, Ma, Gy  & SVM      &  10       & \cmark & \xmark& \xmark &   \cmark  & \cc{98}97.95\% &     180s \\

\cite{AminiNPGYK18} &  Free motion &    Ac, Gy  & LSTM      &  47       & \xmark & \cmark& \cmark &   \cmark  & \cc{97}96.7\% &     20s \\

\cite{LiHZ19} &  Free motion &    Ac, Gy  & SVM      &  100       & \err{8}8.33\% & \xmark& \xmark &   \xmark  & \xmark &     5s \\

\cite{AbuhamadAMN20}   &   Free motion &   Ac, Gy, Ma    & LSTM & 84 &  \err{0}0.09\% & 0.96\% & 8.08\% & \cmark &  \cc{98}97.52\% & 0.5s \\

\cite{SongWRX16} &  Eye movement &    Ca        & SVM          &  20        & \err{11}10.61\% & \cmark& \cmark &   \cmark  & \cc{89}88.73\% &     10s \\
\cite{ZhangHXCH17} &  Eye movement &    Ca        & SRC          &  30        & \err{7}6.9\% & \cmark & \cmark &   \cmark  & \cc{93}93.1\% &     130s \\

\midrule

\multicolumn{11}{l}{Ac: Accelerometer, Gy: Gyroscope, Ma: Magnetometer, Pr: Pressure, LMC: Leap Motion Controller, Or: Orientation, RF: Random Forest,}\\
\multicolumn{11}{l}{SVM: Support Vector Machine, DTW: Dynamic Time Wrapping, LSTM: Long Short Term Memory, MLP: Multilayer Perceptron, } \\
\multicolumn{11}{l}{CC: Cross-correlation, k-NN: k-Nearest Neighbor, I-F: Isolation Forest, SRC: Sparse Representation Classification.}\\

\bottomrule
\end{tabular}}
\end{table*}

\BfPara{Smartphone Hardware and Software Capabilities}
The rapid advancements in mobile technologies have increased the performance of smartphones by multiple folds in the recent years. 
The computational capabilities of mobile devices, including multi-core processors, GPUs, and Gigabytes of memory, are comparable to those of normal-use desktop computers.  
Hardware acceleration units, that are available on most smartphones' chipset platforms, \eg Qualcomm, HiSilicon, MediaTek, and Samsung,  
have enabled smartphones to run sophisticated applications that go far beyond standard and built-in phone functions. Moreover, today's smartphones are equipped with a variety of sensors, \eg motion sensors, environmental sensors, and position sensors, that can provide an accurate usage profiling for enhanced user experience.
While standard applications are no longer a challenge with such capabilities, there are many performance requirements and challenges related to adopting continuous behavioral-based authentication on smartphones, especially when using machine learning approaches. 
Such challenges include the following. \cib{1} \textbf{OS-related Development Tools:} The availability of such tools to access and take advantage of the embedded processing acceleration units plays a key role in developing continuous authentication methods.  
Most the of surveyed systems are implemented on Android-based platforms for the ease of access to a variety of developing tools. Studying the effects of the running OS system on obtaining and analyzing behavioral biometrics for continuous authentication is an interesting direction for a future work that is out of scope of this study. \cib{2} \textbf{Machine Learning-based Authentication:} While the current computational and memory power of smartphones allow for model inference, the enrolment phase can be a challenge and may require a server-side training phase.
Through our survey of behavioral-based continuous authentication methods using different modalities, we highlight insights and challenges to advance the application of the addressed modality.  
We note that an efficient implementation of behavioral biometric-based authentication method should account for hardware- and software-independent operation and network connectivity differences to allow for successful system adoption \cite{clarke2007advanced}.

\BfPara{Built-in Methods} Most of the built-in authentication methods are intended for Point-of-Entry level, as continuous implicit authentication is still evolving to meet a specific level of standards.  
To the best of our knowledge, and based on our survey, there has not been any commercial offering of a dedicated built-in continuous authentication method in customer-grade smartphones, making the development of such methods a possible gap to fill with research and development. We note the barrier to the  mass production of built-in authentication capabilities in smartphones is that they need to meet a high standard of security (\eg FAR of 0.01\% in the European Union), which is not met by the current technology. 
In our survey, we highlight variety of challenges that can be pursued to improve the current methods to rise to this level of standards. As standards are clearly outlined for Point-of-Entry authentication, there is still a lack of guidelines for adopting continuous behavioral biometrics as an integral component of the smartphone. However, many of the covered methods are applicable as a running application, given today's devices' resources such as sensors, multi-core processors, and GPUs.






\section{Motion-based Authentication}\label{sec:motion}
Most of today's mobile devices are equipped with motion sensors such as accelerometers and gyroscopes, which can be a valid source for modeling users' behavior. The accelerometer provides the gravitational acceleration in three spatial dimensions (axes), $x$, $y$, and $z$, measured in meter per second squared, where the axes denote the vertical, and left-to-right dimensions \cite{ferrero2015gait}. The gyroscope measures the angular rotation in three dimensions, $x$, $y$, and $z$, in radians per second along the axes \cite{fantana2015movement}. Such sensory data provides a feature space that enables the modeling of users' movement and usage; therefore, a variety of methods revolve around utilizing such data for authentication and security.

Early exploitation of motion sensors includes air-written signatures \cite{laghari2016biometric,casanova2010real} for which the user holds the device and performs an air-written signature as the application is running and recording the user's motion.  Traditionally, signatures are well-known behavioral biometric commonly used for conducting official or commercial transactions \cite{jain2004introduction,kartik2008multimodal,bhattacharya2013offline}. 
However, air-written signatures, while providing a valid method for user authentication, they operate as a point-of-entry authentication and fail to offer covert, transparent, or continuous authentication.
Laghari \etal~\cite{laghari2016biometric} showed that a motion-based signature had achieved a 1.46\% FAR and 6.87\% FRR when tested on a dataset collected from motion sensors of ten participants' smartphones.
While such methods are robust against shoulder surfing attacks \cite{sahami2012assessing}, they \cib{1} require the user input and engagement once authentication is required, \cib{2} fail to offer a continuous transparent authentication, and \cib{3} are secret- and knowledge-based since the user must memorize the used signature.
Similar implementations include waving gestures \cite{hong2015waving}, free-form gestures \cite{fantana2015movement}, and ``picking-up'' movement (\ie picking the phone and raising it for answering a call) \cite{feng2013investigating}.

Ehatisham \etal \cite{EhatishamANAL18} proposed a continuous authentication system that identifies mobile users based on their activity patterns using embedded sensors, \ie accelerometer, gyroscope, and magnetometer. The authors reported an analysis of the system performance when the smartphone is placed at five different locations on the user's body.
Amini \etal~\cite{AminiNPGYK18} introduced {\em DeepAuth}, an LSTM-based user authentication method, which uses sensory data extracted from the accelerometer and gyroscope to model users' behavioral patterns. The experiments, which were carried out on data collected from 47 users with 10--13 minutes each, have shown an average accuracy of 96.7\% for 20 seconds authentication window.
Zhu \etal \cite{ZhuHCL17} introduced a technique based on users' phone-skating behavior captured by motion sensors. The experiments reported an average EER of 1.2\% using data of 20 users.
Lee \etal \cite{LeeL15} introduced an SVM-based system for user authentication using readings from three motion sensors to achieve an average accuracy of 90\% when using data collected from four participants. 

Exploring the effects of using different sensory data augmentation process, Li \etal \cite{LiHZ19} examined five data augmentation methods to authenticate users with {\em SensorAuth}. The overall results of {\em SensorAuth} have shown an EER of 4.66\% when using 5 seconds window. 

Using different motion-based modality,  Zhang \etal \cite{ZhangHXCH17} introduced an eye movement-based implicit authentication method based on eye movement in response to visual stimuli when using a VR headset. The authors reported imposters' detection accuracy of 91.2\% within 130 seconds. 
Song \etal \cite{SongWRX16} conducted a similar study on smartphones to track individual eye movement with the built-in front camera to investigate using gaze patterns for user authentication \cite{SongWRX16}. The authors reported an average system accuracy of 88.73\% when tracking users' eye movement for 10 seconds.

The summary of the related work associated with motion-based user authentication is listed in Table~\ref{table:motion}. 
In this table, the performance metrics and authentication time are reported based on the original referenced paper. We follow this approach for all of the following tables.
Most of the studies use embedded motion sensors such as accelerometer, gyroscope, and orientation sensors. Using motion-based methods for user authentication allowed an authentication accuracy of up to 99.13\% using SVM trained on sensory data collected from motion sensors~\cite{wu2015implicit}.

\begin{table*}[t]
\centering
\caption{Summary of the related work for gait-based user authentication. Each work is identified by the used modalities, utilized sensors, dataset, modeling algorithm, and their performance.}
\label{table:gait}
\scalebox{1}{\begin{tabular}{lllccccccccc}
\toprule
Study & Modalities & Sensors &  Methods & \begin{tabular}[c]{@{}c@{}}\# \\ Users\end{tabular} & EER & FAR & FRR & TPR &  Accuracy & \begin{tabular}[c]{@{}c@{}}Auth. \\ Time\end{tabular} \\
\midrule

\cite{wang2004fusion}           &      Gait   &    Camera              & k-NN           &  20        & \err{9}3.54\% & \cmark & \cmark &   \xmark    & \cc{88}87.5\% &   \xmark \\

\cite{gafurov2009gait}           &      Gait   &    MRC -- ankle              & k-NN           &  21        & \err{5}5\% & \cmark & \cmark &   \xmark    & \cc{86}85.7\% &   \xmark \\

        &       &    MRC -- hip    & k-NN   &  100       & \err{13}13\% & \cmark & \cmark &   \xmark    & \cc{73}73.2\% &   \xmark \\
        &       &    MRC -- pocket    & k-NN   &  50       & \err{7}7.3\% & \cmark & \cmark &   \xmark    & \cc{86}86.3\% &   \xmark \\
        &       &    MRC -- arm    & k-NN   &  30       & \err{10}10\% & \cmark & \cmark &   \xmark    & \cc{72}71.7\% &   \xmark \\


\cite{qian2008people}           &      Gait   &    FSR              & FLD           &  10        & \xmark & 5.07\% & \xmark &   \xmark    & \cc{89}88.8\% &   0.127s \\

\cite{ferrero2015gait}           &      Gait   &    Ac              & Guidelines          &  \xmark        & \xmark & \xmark & \xmark &   \xmark   & \xmark&   \xmark \\

\cite{thang2012gait}           &      Gait   &    Ac              & SVM           &  11        & \xmark & \xmark & \xmark &   \xmark   & \cc{93}92.7\% &   \xmark \\

\cite{hoang2013adaptive}      &      Gait   &    Ac              & SVM           &  14        & \xmark & \xmark & \xmark &   \xmark   &  \cc{92}$\text{91.33}\pm \text{0.67}$\% &   \xmark \\

\cite{mantyjarvi2005identifying}      &      Gait   &    Ac              & CC           &  36        & \err{7}7\% & \cmark & \cmark &   \xmark    &  \cmark &   \xmark \\

\cite{muaaz2013analysis}      &      Gait   &    Ac              & DTW-SVM           &  51        & \err{33}33.3\% & \cmark & \cmark &   \cmark    &  \cc{53}53\% &   \xmark \\

\cite{nickel2011scenario}      &      Gait   &    Ac              & CRM           & 48       & \err{22}21.7\% & \cmark & \cmark &   \cmark    &  \cc{53}53\% &   30s \\

\cite{nickel2012authentication}      &      Gait   &    Ac              & k-NN           & 36       & \err{8}8.24\% & \xmark & \xmark &   \xmark   &  \xmark &   1.7m \\

\cite{hoang2015gait}      &      Gait   &    Ac              & FC         & 38      & \err{4}3.5\% & 0 & 16.18\% &   \xmark    &  \xmark &  \xmark \\

\cite{vildjiounaite2006unobtrusive}      &      Gait   &    Ac--In-hand      & CC         & 31      & \err{17}17.2\% & \xmark & \xmark &   \xmark    &  \xmark &  \xmark \\

 &     &    Ac--Chest      & CC         &      & \err{15}14.8\% & \xmark & \xmark &   \xmark    &  \xmark &  \xmark \\

 &     &    Ac--Hip      & CC         &      & \err{14}14.1\% & \xmark & \xmark &   \xmark    &  \xmark &  \xmark \\

 &     &    Ac--In-hand      & FFT         &       & \err{14}14.3\% & \xmark & \xmark &   \xmark    &  \xmark &  \xmark \\

 &     &    Ac--Chest      & FFT         &      & \err{14}13.7\% & \xmark & \xmark &   \xmark    &  \xmark &  \xmark \\

 &     &    Ac--Hip      & FFT         &      & \err{17}16.8\% & \xmark & \xmark &   \xmark    &  \xmark &  \xmark \\


\cite{nickel2013classifying}      &      Gait   &    Ac              & HMM         & 48      & \err{6}6.15\% & \cmark & \cmark &   \cmark    &  \xmark &  33s \\

\cite{xu2018keh}      &      Gait   &    Ac, Gy, Co, PEH, EEH              & PMSSRC         & 20      & \err{12}6--12.1\% & \cmark & \cmark &   \cmark    &  \cc{96}96\% &  1.6ms \\

\cite{kolokas2019gait}      &      Gait   &    Ac, Gy, Camera    & Matching         & 10      & \xmark & \cmark & \cmark &   \cmark    &  \cc{91}91\% &  15-75s \\

     &         &       &          & 20      & \err{21}20.8\% & \cmark & \cmark &   \cmark    &  \cc{81}81.3\% &  \\
     &         &       &          & 30      & \xmark & \cmark & \cmark &   \cmark    & \xmark &  \\


\cite{ferreira2017user}      &      Gait   &    Ac, Gy, Ma              & SVM \& RF         & 50      &  \xmark & \xmark & \xmark &   \xmark   &  \cmark &  6.4s \\

\cite{sun2018artificial}      &      Gait   &    Ac, Gy, Ma              & CC--FC         & 15      &  \err{6}5.5\% & \cmark & \cmark &   \xmark   &  \cc{95}95\% &  12s \\

\midrule

\multicolumn{11}{l}{MRC: Cyclic Rotation Metric, FSR: Force Sensing Resistor, Ac: Accelerometer, Gy: Gyroscope, Co: Compass, PEH: Piezoelectric Energy Harvester,}\\
\multicolumn{11}{l}{EEH: Electromagnetic Energy Harvester, Ma: Magnetometer, SVM: Support Vector Machine, RF: Random Forest, DTW: Dynamic Time Wrapping,} \\
\multicolumn{11}{l}{CC: Cross-correlation, k-NN: k-Nearest Neighbor, FLD: Fisher Linear Discriminant, CRM: Cyclic Rotation Metric, FC: Fuzzy Commitment,}\\
\multicolumn{11}{l}{FFT: Fast Fourier Transform, HMM: Hidden Markov Model, PMSSRC: Probability-based Multi-Step Sparse Representation Classification.}\\

\bottomrule
\end{tabular}}
\end{table*}

\BfPara{Insights and Challenges}
While motion-based user authentication methods can detect and classify legitimate users, it has been shown that using the motion-based authentication alone achieves a relatively lower accuracy (up to 96.87\%) in comparison with methods that incorporate multiple modalities.
For example, using the keystroke dynamics along with motion sensors, \ie as an indication of an active usage of the device, enables a higher authentication accuracy \cite{wu2015implicit}. 
Note that some motion-based modalities, \eg waving gestures, free-form gestures, motion-based signature, in-air writing, fail to offer
a covert continuous authentication. Therefore, numerous studies have explored other modalities that rely on behavioral biometrics captured by the motion sensors and wearable devices to implement a transparent continuous authentication.
Handling information from multiple sensors and sources, \eg wearable devices, for an implicit authentication is a challenging task that requires several on-device data preprocessing techniques, temporal data alignment, and accurate modeling and matching.

Common open challenges of using motion-based continuous authentication on smartphones include the following.
\cib{1} \textbf{Power Consumption:} Intuitively, continuous authentication schemes, in general, consume power. This consumption is due to multiple processing components of the adopted method, data collection and sampling, feature extraction, model inference, and matching algorithms. For example, a study by Lee \etal \cite{LeeL17} shows that continuously querying of sensors data at 50Hz sampling rate  for 12 hours can consume up to 5\% of the battery life even without active usage (\ie the device is locked). Using a higher sampling rate can result in significantly higher power consumption \cite{LeeL17, ZhuQXZS19}. Note that power consumption varies from a device to another, considering the hardware configurations and processing units. For example, a study by \cite{ZhuQXZS19} shows that the power consumption of running {\em RiskCog} for three hours with a 50Hz data sampling rate on three devices as follows: Samsung N9100 (4.4\%), Sony Xperia Z2 (3.6\%), and MI 4 (4.2\%).
\cib{2} \textbf{Computation and Memory Overhead:} Motion-based continuous authentication requires continuous collection and processing of data as well as high-frequency authentication via model or matching algorithm inference. Moreover, data records within the collection timeframe and predefined operational thresholds increases the memory overhead. Optimizing the computational and memory requirements for motion-based schemes is considered an open challenge.
\cib{3} \textbf{Adversarial Attacks:} 
Motion-based authentication schemes can be vulnerable to attacks, including observation-based attacks (\eg observing and reproducing in-air handwriting and gestures), and sensor-based inference attacks (\eg sensor-based side-channel inference attacks). While behavioral biometrics can be accurately captured by sensors, sensors data can be collected by a variety of applications that may present a threat to the adopted modality. Addressing such attacks is an interesting and an open research direction.

\begin{table*}[t]
\centering
\caption{Summary of the related work for keystroke dynamics-based user authentication. Each work is identified by the used modalities, utilized sensors, dataset, modeling algorithm, and their performance.}
\label{table:keystroke}
\scalebox{1}{\begin{tabular}{lllccccccccc}
\toprule
Study & Modalities & Sensors &  Methods & \begin{tabular}[c]{@{}c@{}}\# \\ Users\end{tabular} & EER & FAR & FRR & TPR   & Accuracy & \begin{tabular}[c]{@{}c@{}}Auth. \\ Time\end{tabular} \\
\midrule

\cite{monrose2000keystroke}   &  Keystroke Dynamics   &    NA      & k-NN   & 63  &  \xmark & \xmark & \xmark &   \xmark    & \cc{85}83.22--92.14\% &  \xmark \\

\cite{joyce1990identity}   &  Keystroke Dynamics   &    NA      & Matching   & 33  &  \xmark & 0.25\% & 16.36\% &   \xmark    & \xmark &  \xmark \\

\cite{mcloughlin2009keypress}   &  Keystroke Dynamics   &    NA      & Distance \& CI   & 3  &  \xmark & \cmark & \cmark &   \xmark     & \xmark &  \xmark \\

\begin{tabular}[c]{@{}c@{}}\cite{zahid2009keystroke} \\\\ \\ \\ \\ \end{tabular}   &  \begin{tabular}[c]{@{}c@{}}Keystroke Dynamics \\ \\ \\\\ \\ \end{tabular}     &    \begin{tabular}[c]{@{}c@{}}NA           \\ \\ \\ \\ \\ \end{tabular}    & \begin{tabular}[c]{@{}c@{}}RBFN \\Fuzzy \\PSO-Fuzzy \\GA-Fuzzy\\ PSO-GA Fuzzy\end{tabular}   & \begin{tabular}[c]{@{}c@{}}25            \\25 \\25 \\25 \\25 \end{tabular}  &  
\begin{tabular}[c]{@{}c@{}}\xmark  \\\xmark \\\xmark \\\xmark \\\xmark \end{tabular} & 
\begin{tabular}[c]{@{}c@{}}36\%   \\ 18.6\%\\8.09\% \\8.79\% \\2.07\% \end{tabular} & 
\begin{tabular}[c]{@{}c@{}}26.6\%  \\19\% \\ 7.58\%\\ 7.94\% \\ 1.73\%\end{tabular} &   
\begin{tabular}[c]{@{}c@{}}\xmark  \\\xmark \\\xmark \\\xmark \\\xmark \end{tabular}& 
\begin{tabular}[c]{@{}c@{}}\xmark  \\\xmark \\\xmark \\\xmark \\\xmark \end{tabular} &  
\begin{tabular}[c]{@{}c@{}}\xmark  \\\xmark \\\xmark \\\xmark \\\xmark \end{tabular} \\

\cite{urtiga2011keystroke}   &  Keystroke Dynamics   &    NA      & Distance   & 15  &  \xmark & 12.97\% & 2.25\% &   \xmark     & \xmark &  \xmark \\

\cite{hwang2009keystroke}   &  Keystroke Dynamics   &    NA      & Distance   & 25  &  \err{4}4\% & \xmark & \xmark &   \xmark     & \xmark &  632-2151ms \\

\cite{wu2015smartphone}   &  Keystroke Dynamics   &    NA      & SVM   & 10  &  \xmark & \xmark & \xmark &   98.7\%     & \cc{99}98.6\% &  \xmark \\

\cite{giuffrida2014sensed}   &  Keystroke Dynamics   &    Ac, Gy      & k-NN   & 20  & \err{0} 0.08\% & \xmark & \xmark &   \xmark   & \xmark & 200ms \\

\cite{inguanez2016securing}   &  Keystroke Dynamics   &    NA      & MLP   & 32  &  \xmark & 6.33\% & 4.89\% &  95.11\%   & \cc{95}94.81\% & \xmark \\

\cite{cilia2018multi}   &  Keystroke Dynamics   &    NA      & SVM   & 24  & \err{2} 1.42\% & 2\% & 1\% &  99\%   & \cc{99}99\% & \xmark \\

\cite{anusas2019strengthening}   &  Keystroke Dynamics   &    Ac      & SVM   & 5  & \err{5} 5.1\% & \xmark & \xmark &  \xmark  & \cc{98}97.9\% & \xmark \\

\cite{shankar2019intelligent}   &  Keystroke Dynamics   &    NA      & DAE-SR   & 10  &  \err{5}5\% & \xmark & \xmark &  91.8\%  & \cc{95}95\% & \xmark \\

\cite{DraffinZZ13}   &  Keystroke Dynamics   &    NA      & MLP   & 13  &  \xmark & 14\% & 2.2\% &  \xmark  & \cc{98}86\% & \xmark \\

\cite{Mondal2017}   &  Keystroke Dynamics   &    NA      & MCF   & 64  &  \xmark & \xmark & \xmark &  \xmark  & \cc{90}89.7\% & \xmark \\

\midrule

\multicolumn{11}{l}{Ac: Accelerometer, Gy: Gyroscope, CI: Confidence Interval, RBFN: Radial Basis Function Network, PSO: Particle Swarm Optimization,}\\
\multicolumn{11}{l}{GA: Genetic Algorithm, DAE-SR: Deep Auto Encoder and Softmax Regression, MCF: Multi-Classifier Fusion, SVM: Support Vector Machine,} \\
\multicolumn{11}{l}{k-NN: k-Nearest Neighbor, MLP: Multilayer Perceptron.}\\

\bottomrule
\end{tabular}}
\end{table*}

\section{Gait-based Authentication}\label{sec:gait}
Gait recognition has gained increased interest in recent years, especially with the vast adoption of mobile and wearable sensors. Gait recognition is defined as the process of identifying an individual by the manner of walking using computer vision and/or sensory data collected from environmental and wearable sensors \cite{derawi2010unobtrusive}. Computer vision approaches for gait recognition include segmenting the individual's images while walking and capturing the features that enable accurate recognition \cite{wang2004fusion}. While using sensory data, including \cib{1} adopting floor senors where the gait-related features are captured once the person walks on them \cite{gafurov2009gait,qian2008people}, \cib{2} adopting wearable sensors that aims to collect information that enables gait recognition \cite{gafurov2009gait}. For mobile security and authentication, gait recognition is usually done using wearable sensors, especially the reading of the motion sensors (\eg accelerometer) of the mobile device, to enable continuous transparent authentication.

The general approach to gait recognition includes four steps, \cib{1} data acquisition step in which the device is placed in a certain way that enables the walk activity recording, 
\cib{2} data preprocessing step for reducing the introduced noise by the data collection method or other environmental factors, 
\cib{3} walk detection step using either traditional cycle or machine learning techniques, and \cib{4} analysis step \cite{ferrero2015gait}. 
Handling the data acquisition process requires accurate readings from motion sensors as the user places the device in a predefined manner such as carrying the device inside of a pouch \cite{nickel2012authentication}, in the pants pocket \cite{ferrero2015gait,hoang2015gait}, or in hand \cite{vildjiounaite2006unobtrusive}. 
Studies conducted for mobile security using gait-based biometrics usually include data collection from a population of size equal to or less than 50 participants \cite{nickel2012authentication,hoang2015gait,vildjiounaite2006unobtrusive}, and processed in controlled conditions to minimize the effects of outside factors \cite{nickel2013classifying}. Even though some studies have attempted to capture gait-related metrics from a real-world collection of sensory data, such as the study by Nickel and Busch \cite{nickel2013classifying}, generally, the data collection requires an ideal setting at least in one aspect (\eg walking patterns or floor condition) \cite{neal2016surveying}.

The second step after acquiring the data, the preprocessing step takes place to clean, reduce the noise, and normalize the data. The major task in this regard is the noise reduction considering various possible noise sources, such as environmental and gravitational factors, floor conditions, and the users' shoes or other wearable materials. Since the gait-related features rely heavily on readings from motion sensors, such as the accelerometer, which are very sensitive, the adopted method should account for further noise \cite{hoang2015gait}. Such noises can be handled using linear interpolation and filtering techniques, while environmental noise adds much complexity to the walk detection task, which can be minimized using activity recognition to remove any irrelevant data  \cite{nickel2012authentication}. 
For the walk detection, cycles (\ie the time between two paces bounded by maximum and minimum threshold across the three axes) or machine learning techniques are both utilized in the literature.
Cycle-based approaches are commonly used since the average cycle length is easily and simply calculated to detect cycles by moving forward or backward in intervals of the average cycle length with some correction measurement. On the other hand, machine learning-based approaches have shown to be accurate for automatic walk detection \cite{nickel2013classifying}. Such techniques require readings of the sensory data, preprocessing phase to reduce the noise and normalize the data, and a walk detection model that leverages the lowest and highest values for thresholding and the decision. 

The final step of gait recognition is the analysis of the time intervals, frequencies, or both. 
Using time intervals analysis, some metrics can be extracted and studied, such as cycle statistics, including the minimum, average, maximum acceleration values, and cycle lengths and frequencies.
Moreover, cycle variance and stability are measured by acceleration moments  \cite{ferrero2015gait,derawi2010unobtrusive}. 
Using frequency analysis, usually conducted using Discrete or Fast Fourier Transforms, it has been shown that the first few coefficients resulting from each conversion are highly relevant for detecting distinctive gait patterns \cite{ferrero2015gait}.

Wang~\etal~\cite{wang2004fusion} and Gafurov~\etal~\cite{gafurov2009gait} used a k-NN model to classify legitimate users using gait-based features, where Wang~\etal uses the camera to capture the user movement, and Gafurov~\etal captures the user movement using cyclic rotation metric device attached to different places of the body (ankle, hip, pocket, arm). Both studies achieved an accuracy of above 85\%, with EER of 3.54\% and 5\%, respectively. Multiple studies used accelerometer as a standalone sensor to capture user movement for user authentication task~\cite{ferrero2015gait,thang2012gait,hoang2013adaptive,mantyjarvi2005identifying,muaaz2013analysis,nickel2011scenario,nickel2012authentication,hoang2015gait,vildjiounaite2006unobtrusive,nickel2013classifying}. 

Both Thang~\etal~\cite{thang2012gait} and Hoang~\etal~\cite{hoang2013adaptive} collected data of 11-14 users and used SVM-based models for capturing user patterns, achieving a nearly the same accuracy of 92\%. In addition, Hoang~\etal~\cite{hoang2015gait} achieved an EER of 3.5\% by using a fuzzy commitment algorithm on a study sample of 38 users, outperforming its counterparts. Others~\cite{xu2018keh,kolokas2019gait,ferreira2017user,sun2018artificial} incorporated different sensors to capture the motion aspects of the users, achieving an accuracy of up to 96\% by using accelerometer, gyroscope, compass, piezoelectric energy harvester, and electromagnetic energy harvester~\cite{xu2018keh}.
The summary of the gait-based user authentication methods is shown in Table~\ref{table:gait}. 

\BfPara{Insights and Challenges}
Similar to motion-based user authentication methods, gait-based methods do not achieve a high relative accuracy nor precision in user authentication tasks. 
Generally, gait-based user authentication methods are feasible in specific applications, which requires capturing the user's gait traits while moving, \eg player detection in a team-based sport via wearable sensing devices. 
Applying gait-based authentication for smartphone users requires addressing a variety of challenges, such as the following. \cib{1} \textbf{Data Sources:} Collecting gait-related sensory data requires visual information as well as motion information from multiple sensors. \cib{2} \textbf{Sensors Placement:} As changing the placement of the device can significantly change the sensory readings. \cib{3} \textbf{Adopting Alternatives:} As gait-based authentication fails to provide a continuous authentication when the user is  not moving. \cib{4} \textbf{Usability:} As the user state at the enrollment stage may differ from the state the inference stage. Moreover, the gait-based traits are highly dependent on the user's physical state when capturing the data. Such challenges may explain the relatively low accuracy of the gait-based authentication methods.

\section{Keystroke-based Authentication}\label{sec:keystroke}
One of the earliest behavioral authentication methods is based on studying the keystroke dynamics. Most keystroke dynamics-based methods are cost-effective and do not require additional modules to operate \cite{karnan2011biometric}. 
During the usage of the device, when a key input is required (\eg texting), the keystroke dynamics-based authentication method continuously validates the user since behavioral dynamics can be distinctive across users. 
Conducting authentication via keystroke dynamics requires analyzing and capturing the distinctive features and patterns of users' keystrokes when using the device \cite{teh2013survey,6496441}. Common features include:
\cib{1} Keypress frequency, which calculates the frequency of keypress events.
\cib{2} Key release frequency, which calculates the frequency of key release events.
\cib{3} Latency and hold time, which calculates the rates of press-to-press, press-to-release (which is also known as the hold time), release-to-release, and release-to-press events.\cib{4} Finger's pressure while touching the screen.
\cib{5} Pressed area size by the user's fingers.
\cib{6} Error rate, which is the frequency of using backspaces or deletion option.

Using keystroke dynamics for authentication or user validation has been adopted on traditional computers before their application to smartphones \cite{monrose2000keystroke}. 
Even though it seems to be an easier task to implement a keystroke dynamics-based authentication on computers due to the less complex feature space, Joyce and Gupta \cite{joyce1990identity} showed the uniqueness of both written signatures and typing behavior are originated from the physiology of the neurological system.

Recent application of keystroke dynamics takes advantage of embedded sensors (\eg motion sensor on smartphones), to improve the authentication accuracy, especially when there the key-based input is unavailable \cite{stanciu2016effectiveness,sitova2015hmog}. 
Another distinction between applying keystroke dynamics-based methods on smartphones and computers is the large space of key-based input in the smartphone since it includes touches and swipes that meant for interacting with the applications without typing textual content \cite{mcloughlin2009keypress}. 
Several studies have addressed the generalization of these methods to different types of input. For instance, McLoughlin \etal~\cite{mcloughlin2009keypress} showed that using key press and release frequencies and the latency between two presses contribute greatly to establishing distinctive keystroke behavior for users. The authors showed that the application should account for the inconsistencies in recorded data by introducing weights based on the variance of data (\ie lower variance gets higher weights). Their results show an accuracy of more than 90\%, establishing the validity of using keystroke dynamics as a biometric for authentication with minimal computational overhead and increased usability.

Buriro \etal \cite{BuriroCGF18} designed an authentication scheme based on the user's hand movements and timing features as they enter ten keystrokes. The authors conducted experiments using data collected from 97 participants and reported an authentication accuracy of 85.77\% and FAR of 7.32\%. 
Similarly, Zahid \etal~\cite{zahid2009keystroke} studied keystroke behavior of 25 users, including features such as the hold time, error rate, and latency. 
The authors suggested a fuzzy classifier to account for the diffused features space and argued that presenting the classification task of keystroke behavior as an optimization problem benefits the robustness of the model when compared to similarity-based methods \cite{urtiga2011keystroke}.
Using a fuzzy classifier with Particle Swarm Optimization and Genetic Algorithms, their proposed method showed 0\% FRR and 2\% FAR, suggesting high security and usability potential. 
However, keystroke dynamics are often incorporated with other modalities for improving performance and accuracy. For instance, Hwang \etal~\cite{hwang2009keystroke} suggested including rhythm and tempo as components for studying keystroke dynamics, \ie a user is required to follow a distinct and consistent timing pattern for accurate keystroke-based authentication. For example, a given term can be entered digit by digit separated with subsequent short and long pauses that are controlled by tempo cues, \eg a metronome for counting pause intervals. In their study, the authors showed an average improvement of about 4\% in the EER evaluation metric when using artificial rhythmic input with tempo cues in comparison to natural rhythms. However, adopting such methods adds complexity to the usability aspect.

Using smartphone embedded sensors to support keystroke dynamics-based authentication has been repeatedly suggested to improve the performance and to provide transparent authentication.
\cite{wu2015smartphone} proposed incorporating velocity-related metrics to reach an accuracy of 98.6\% for classifying data from ten users using an SVM classifier. Similarly, Giuffrida \etal~\cite{giuffrida2014sensed} proposed incorporating keystroke data with motion sensors data, namely, accelerometer and gyroscope, to conclude that metrics obtained from the accelerometer data are more useful than those obtained from the gyroscope. The authors showed that combining features from motion sensors with keystroke metrics provides similar results as adopting only the motion sensors-related features alone, \ie the study shows that sensor-related features can be more useful than keystroke dynamics in terms of authentication. However, obtaining and analyzing high-frequency sensory data can be power consuming. 
Table~\ref{table:keystroke} shows a list of authentication methods based on keystroke dynamics. The proposed approaches show a promising direction for using this modality for user authentication, achieving an accuracy of up to 99\% by Cilia~\etal~\cite{cilia2018multi}.

\BfPara{Insights and Challenges}
Keystroke dynamics-based methods have several advantages, such as (a) their high authentication accuracy that can reach up to 99\%, (b) high power-efficiency in comparison with other methods, and (c) hardware independence, since these methods can operate with either physical or on-screen keyboards.
However, implementing a keystroke dynamics-based approach can be challenging for several reasons. \cib{1} \textbf{User Behavioral Changes:} 
Capturing keystroke dynamics as a behavioral modality under uncontrolled conditions, \eg user's activity (standing, walking, \etc), user's emotional or physical state change, and the in-use application, is challenging and requires testing under these non-trivial scenarios. 
\cib{2} \textbf{Feature Extraction and Selection:} The extracted metrics should be robust against noise and behavioral changes. Considering the limited space of features, recent studies have considered incorporating other modalities to extend the feature space, thus allowing for the selection of a distinctive user representation that can be generalized to a relatively large population.
\cib{3} \textbf{Adopting Alternatives:} Since these methods operate only
when the user interacts with the keyboard, the implicit authentication module should allow for possible alternatives when the user uses the device without typing (\eg watching a video, placing a call, \etc).
Other challenges can be related to typing with different languages and whether the user's typing behavior changes across languages, which require further attention through further research.

\begin{table*}[t]
\centering
\caption{Summary of the related work for gesture-based user authentication. Each work is identified by the used modalities, utilized sensors, dataset, modeling algorithm, and their performance.}
\label{table:gesture}
\scalebox{1}{\begin{tabular}{lllccccccccc}
\toprule
Study & Modalities & Sensors &  Methods & \begin{tabular}[c]{@{}c@{}}\# \\ Users\end{tabular} & EER & FAR & FRR & TPR &  Accuracy & \begin{tabular}[c]{@{}c@{}}Auth. \\ Time\end{tabular} \\
\midrule

\cite{mondal2015swipe}   &  Swipe Gesture   &       NA   & ANN-CPANN   & 71  &   \xmark & 0.08\% &  0  &\xmark &  \xmark & \xmark \\

\cite{nohara2016personal}   &  Flick Gesture   &      AC, Gy    & SOM   & NA &   \xmark & \xmark &  \xmark &\xmark & \cc{93}92.8\% & \xmark \\

\cite{lin2012new}   &  Flick Gesture    &      Or    & k-NN   & 16 &   \err{7}6.85\% & \cmark &  \cmark &\xmark & \xmark  & $<$100ms \\

\cite{lu2015safeguard}   &   Slide Gesture  &      NA    & SVM  & 60 &  \err{0}0.01--0.02\%  & 0.03\% &  0.05\% &\xmark &  \xmark & 0.3s \\

\begin{tabular}[c]{@{}c@{}}\cite{jain2015exploring} \\ \\ \end{tabular}  &  \begin{tabular}[c]{@{}c@{}}Swipe Gesture  \\\\  \end{tabular}   &     \begin{tabular}[c]{@{}c@{}}Ac, Or \\\\  \end{tabular}    & \begin{tabular}[c]{@{}c@{}}MHD  \\ DTW \end{tabular}  & \begin{tabular}[c]{@{}c@{}}104  \\ 104 \end{tabular}  &  \begin{tabular}[c]{@{}c@{}}\err{0}0.31\%  \\\err{1} 1.55\% \end{tabular}  & \begin{tabular}[c]{@{}c@{}}\cmark  \\ \cmark \end{tabular}  &  \begin{tabular}[c]{@{}c@{}}\xmark  \\ \xmark \end{tabular}  & \begin{tabular}[c]{@{}c@{}}\cmark  \\ \cmark \end{tabular} & \begin{tabular}[c]{@{}c@{}}\xmark  \\ \xmark \end{tabular} & \begin{tabular}[c]{@{}c@{}}\xmark  \\ \xmark \end{tabular} \\

\cite{shih2015flick}   &   Flick Gesture  &      Ac    & Na\"ive Bayes  & 10 &  \xmark  & 1.3\% &  8\% &92\% &  \cc{98}98\% & \xmark \\

\cite{saevanee2008user}   &   Touch \& keystroke &      NA    & k-NN  & 10 & \err{1} 1\%  & \xmark &  \xmark & \xmark &  \cc{99}99\% & 20ms \\

\cite{xu2014towards}   &   Keystrokes/Touch/Handwriting &      NA    & SVM-RBF  & 32 &  \err{9}0.75--8.67\%  & \xmark &  \xmark & \xmark &  \cmark & \xmark\\

\cite{nixon2016slowmo}   &   Gesture &      NA    & MGGM  & 20 & \cmark  & \xmark &  \xmark & \xmark &   \cc{89}89\% & 53ms\\

\cite{nader2015designing}   &   Gesture &      NA    & PSO-RBFN  & 20 & \err{8}8.1\%  & 2\% &  8.2\% & \xmark &  \xmark & \xmark\\

\cite{antal2016biometric}   &   Swipe Gesture &      Or    & RF  & 40 & \err{0}0.2\%  & \xmark &  \xmark & \xmark &  \xmark & \xmark\\

\cite{zaliva2015passive}   &   Touch Gesture &      NA    & RF  & 14 & \xmark  & \xmark &  \xmark & 99.9\% &  \cc{100}99.9\% & 12.6s\\

\cite{primo2015music}   &   Swipe Gesture &      NA    & RF  & 34 & \err{23}16.22--22.94\%  & \xmark &  \xmark &\xmark &  \xmark & \xmark\\


\cite{feng2014tips}   &   Touch Gesture &      NA    & DTW-k-NN  & 23 & \cmark  & \xmark &  \xmark & 91\% &  \xmark & \xmark\\

\cite{shen2015performance}   &   Touch Gesture &      NA    & RF & 71 & \err{2}1.8\%  & 0.1\% &  18.52\% & \xmark &  \xmark & 0.77s\\

\cite{syed2019touch}   &   Touch Gesture &      NA    & RF & 71 & \err{5}5.4\%  & \xmark &  \xmark & \xmark &  \xmark & \xmark\\

\cite{rauen2018gesture}   &   Touch Gesture &      NA    & RF & NA & \xmark  & 2.54\% & 1.98\% & \xmark &  \cc{100}99.68\% & \xmark\\

\cite{rocha2019continuous}   &   Touch Gesture  &      NA    & Matching & 30 & \xmark  & \xmark &\xmark & 93.01\% &  \cc{94}93.76\% & \xmark\\

\midrule

\multicolumn{11}{l}{Ac: Accelerometer, Gy: Gyroscope, Or: Orientation, ANN: Artificial Neural Network, CPANN: Counter Propagation Artificial Neural Network,}\\
\multicolumn{11}{l}{SOM: Self Organizing Maps, k-NN: k-Nearest Neighbor, SVM: Support Vector Machine, MHD: Modified Hausdorff Distance, RF: Random Forest,} \\
\multicolumn{11}{l}{DTW: Dynamic Time Wrapping, RBF: Radial Basis Function, MGGN: Multivariate Gaussian Generative Model, PSO: Particle Swarm Optimization,}\\
\multicolumn{11}{l}{RBFN: Radial Basis Function Network.}\\

\bottomrule
\end{tabular}}
\end{table*}

\section{Touch Gesture-based Authentication} \label{sec:gesture}
Using touch gestures as a biometric modality extend landscape of transparent authentication applications to include a variety of devices with touchscreen unit (\eg smartwatches, digital cameras, navigation systems, and monitors) \cite{neal2016surveying}. 
Several studies have investigated the touch gestures as a behavioral biometrics for continuous authentication since it can be convenient and cost-effective. Touch gestures include swipes \cite{mondal2015swipe,mondal2015continuous}, flicks \cite{nohara2016personal,lin2012new,shih2015flick}, slides \cite{lu2015safeguard}), and handwriting \cite{alsulaiman2008user}.
The distinction between keystroke dynamics and touch gestures can be summarized in the input form for users and the method of input.
The commonalities between the two modalities are the space of improvement when accounting for motion sensors \cite{nohara2016personal,jain2015exploring}.
Therefore, many studies have incorporated motion-based features to gesture-based methods \cite{shih2015flick}. 
Considering features from touch gestures enables accurate authentication with an accuracy reaching to 99\% and minimal EER such as 0.03\% when applying k-Nearest Neighbors classifier or other distance-based classifiers \cite{jain2015exploring}.

Leveraging the abundance of information generated by the operating system of smartphones, a large number of features can be extracted from touch gestures such as the reading from the accelerometer, pressure, gravity, velocity, touch area, and time-related measurements. Such features allow for accurate calculation of the gesture statistics and developing patterns for user authentication \cite{saevanee2008user,xu2014towards,cai2011touchlogger,nixon2016slowmo,nader2015designing}.  Antal \etal~\cite{antal2016biometric} extended the feature space of swipe gestures to include touch duration, trajectory length, acceleration, average speed, touch pressure, touch area, and gravity readings.  Using data from 40 users, including 58 samples, the authors performed one and two-class classification using multiple classifiers such as Bayes Net, k-Nearest Neighbor, and Random Forests. The authors reported that Random Forests showed an EER of  0.004\%. Their results showed that the device motion and positioning are important factors in distinguishing users.

Since touch gestures are commonly known as soft biometrics that could enable the recognition of gender and proportional measures such as physical attributes including hand size, forearm length, and height, they are beneficial in criminal investigations. 
Miguel \etal~\cite{miguel2016predicting} proposed studying the swipe gesture for gender prediction using a variety of features including the swipe's length, width, touch area, pressure, velocity, acceleration, start-to-end angle, and others. The authors showed that applying a multi-linear logistic regression classifier for gender prediction achieves an accuracy of 71\% when the direction of the swipe is down-to-up. Using a fusion of swipe direction-based decision, the accuracy reaches 78\%. 
Similarly, Bevan and Fraser \cite{bevan2016different} investigated the relationship between swipe gestures, thumb length, and gender. 
Using data from 178 users performing one-hand gestures using the thumb, the authors collected 21,360 samples of swipes in various directions. Among the calculated features, the results showed a strong correlation between thumb length and gestures, and they reflected in the velocity, acceleration, and completion time.  Moreover, the study also showed that male users completed the gestures at a higher speed than female users.

The landscape of using touch gestures as behavioral biometrics for user authentication includes devices designed for users with disabilities. 
For example, Azenkot \etal~\cite{azenkot2012passchords} proposed PassChords, which designed for authenticating users with vision impairments using a predefined sequence of screen taps. Another application is proposed by \cite{zaliva2015passive} for users with finger injuries, which uses the finger's trajectory and posture before touching the screen using its positioning and proximity. For this application, the direct touch gesture (\ie the contact with the screen) is not fully required, and only the proximity-related measurement is possibly feasible to authenticate users.

Several studies have shown that gesture-based authentication schemes are application-dependant, and gesture-based data can vary significantly from one application to another, which makes the generalization aspect of gesture-based schemes for continuous authentication across different applications is limited \cite{primo2015music,feng2014tips,shen2015performance,khan2014itus}. Therefore, a ``context-aware'' approach is a potential solution to generalize gesture-based methods. Khan and Hengartner \cite{khan2014towards} showed that the performance of gesture-based methods could be improved by allowing context-aware implementation where different applications control the tuning of features. To this end, the authors used the Kullback-Leibler (KL) divergence metric, which is shown to differ by application indicating the importance of accounting and tuning the features based on the used application. Using data of 32 users who were instructed to use four different applications during the data collection process, the experimental results showed that using the ``context-aware'' approach improves the accuracy of the device-centric approach.

Table~\ref{table:gesture} shows a list of proposed gesture-based authentication methods using varieties of touch gestures and machine learning models. Random forest, in particular, is among the top achieving and adopted models in this modality-based method, with an accuracy above 99\% as shown in~\cite{rauen2018gesture} and \cite{zaliva2015passive}.

\begin{table*}[tb]
\centering
\caption{Summary of the related work for voice-based user authentication. Each work is identified by the used modalities, utilized sensors, dataset, modeling algorithm, and their performance.}
\label{table:voice}
\scalebox{1}{\begin{tabular}{lllccccccccc}
\toprule
Study & Modalities & Sensors &  Methods & \begin{tabular}[c]{@{}c@{}}\# \\ Users\end{tabular} & EER & FAR & FRR & TPR &  Accuracy & \begin{tabular}[c]{@{}c@{}}Auth. \\ Time\end{tabular} \\
\midrule

\cite{miguel2016interaction}   &   Voice  &      Ca, Mi    & Matching & 27 & \xmark  & \xmark &3\% & \xmark &  \cc{93}93\% & $<$ 24.7s\\

\cite{zhang2017hearing}   &   Voice  &      Sp, Mi    & CC & 21 & \err{1}1\%  & 1\% &\xmark & \xmark &  \cc{99}99.34\% & 0.5s\\

\cite{lu2019lip}   &   Voice  &      Sp, Mi    & GMM & 104 & \xmark  & \xmark &\xmark & 99\% &  \cc{95}95\% & \xmark\\

\cite{wang2019voicepop}   &   Voice  &      Mi    & PCA-SVM & 18 & \err{5}5.4\% & 2\% &\xmark & 93\% &  \cc{93}93.5\% & \xmark\\

\cite{yan2016usable}   &   Voice  &      Mi    & DTW & 15 & \xmark & 1\% & 15\% & \xmark &  \cc{89}88.6\% & \xmark\\

\cite{gofman2016hidden}   &   Voice  &      Mi    & HMM & 54 & \err{22}21.58\% & \xmark & \xmark & \cmark & \xmark & 0.07s\\

\cite{kim2008multimodal}   &   Voice  &      Mi    & GMM & 50 & \err{6}6.24\% & \cmark & \cmark & \xmark & \xmark & 10.76s \\

\cite{johnson2013secure}   &   Voice  &      Mi    & GMM & 48 & \err{6}6\% & \cmark & \cmark & \xmark & \xmark & \xmark \\

\cite{zhang2016voicelive}   &   Voice  &      Mi    & Similarity & 12 & \err{1}1.01\% & 1\% & \xmark & 99\% & \cc{99} 99.3\% & \xmark \\

\midrule

\multicolumn{11}{l}{Ca: Camera, Mi: Microphone, Sp: Speaker, GMM: Gaussian Mixed Model, PCA: Principle Component Analysis,}\\
\multicolumn{11}{l}{CC: Cross-Correlation, SVM: Support Vector Machine, DTW: Dynamic Time Wrapping, HMM: Hidden Markov Model. } \\

\bottomrule
\end{tabular}}
\end{table*}

\BfPara{Insights and Challenges}
Similar to keystroke dynamics-based methods, gesture-based authentication methods have several advantages, including (a) their high authentication accuracy, which can reach up to 99.9\%, (b) operating efficiently in terms of both power and computation, (c) conveying high resilience against mimicry attacks since gesture-based modality incorporates multiple independent features, restricting the ability of an impostor to successfully reproduce one feature given another. Moreover, using a high sampling rate (\ie small timeframe) makes it difficult to observe and replicate the touch gestures.
However, several challenges should be considered, including understanding users temporal behavioral changes, applications preferences, users activity, users mobility, \etc



\section{Voice-based Authentication}\label{sec:voice}

Speaker identification using voice-related features has been investigated extensively in the literature \cite{jain2004introduction,tirumala2017speaker}. Voice-related features combine both physiological aspects (\eg vocal tract and lips characteristics) and behavioral traits (\eg emotion- or age-related tones), allowing the speak/voice analysis over large feature space \cite{atal1974effectiveness}. 
Based on \cite{reynolds2002overview}, there are two approaches for using voice to authenticate/identify the speaker, which are as follows. \cib{1} Text-dependent approach, in which users are authenticated based on the matching of speaking a predefined phrase. Since the users speak a certain phrase for authentication, this method is straightforward and very accurate. However, it does not allow for transparent or continuous authentication, and it is not a secret-based method. \cib{2} Text-independent approach in which users are authenticated based on features extracted from the voice regardless of the spoken words. This approach allows higher flexibility, especially in offering transparent authentication, where users are unaware of the service. However, accurate text-independent authentication accuracy faces different challenges due to the dynamic changes in the feature space of voice input accounting for the user condition and other environmental factors.

Speaker recognition using voice features follows the typical pattern recognition system, starting from data collection and preprocessing, going through the feature extraction and selection, and ending with the modeling and pattern recognition. Similar to conventional machine learning-based systems, the quality of features contributes considerably to the accuracy of the speaker recognition. Such features include short-term spectral features, temporal and rhythmic, voice source, prosodic, and conversation-level features \cite{neal2016surveying}. Short-term spectral characteristics represent the resonance attributes of the vocal tract and often extracted with high frequency from 20 to 30 ms timeframes. Prosodic and temporal traits include intonation and rhythmic patterns extracted from long timeframes. Conversation-level features are high-level properties extracted from the textual contents of spoken words, such as word or phrase frequencies.

The quality of features is measured by their distinctive nature and their robustness against possible introduced noise (\eg the user condition and environment) \cite{kinnunen2010overview}. 
In this regard, a study by Reynolds \cite{reynolds2002overview} showed that spectral features provide high-quality, simple, and discriminative feature space.

Using the extracted features, a variety of models are utilized for voice/speaker recognition, such as SVM and Gaussian mixture models \cite{kinnunen2010overview}. Early applications for voice recognition include access control, personalization, and forensic and criminal investigations \cite{reynolds2002overview}. The application landscape has increased to include online banking (\ie conducting a transaction via voice communication as the voice recognition system transparently and continuously authenticate the customer) \cite{neal2016surveying}. 
While voice-based user authentication methods capture the voice using the microphone, different works can be distinguished by the data preprocessing and the utilized machine learning algorithm. Zhang~\etal~\cite{zhang2017hearing} achieved an accuracy of 99.34\% with EER and FAR of 1\% using the cross-correlation method with an authentication time of half a second on a sample size of 21 users. Additionally, using the Gaussian mixed model, Kim~\etal~\cite{kim2008multimodal} and Johnson~\etal~\cite{johnson2013secure} achieved similar EER of around 6\% on a sample size of 50 and 48 respectively. Similarly, Lu~\etal~\cite{lu2019lip} achieved an accuracy of 95\% and TPR of 99\% in conducting user authentication tasks using the Gaussian mixed model with a sample size of 104 users. Multiple machine learning methods may be incorporated for user authentication tasks, Wang~\etal~\cite{wang2019voicepop} used principle components analysis with support vector machine to train data collected from 18 users, achieving an EER of 5.4\% and overall accuracy of 93.5\%. Using a simple approach may outperform powerful machine learning algorithms in user authentication tasks, as Zhang~\etal~\cite{zhang2016voicelive} achieved an accuracy of 99.3\% with EER of 1.01\% and FAR of 1\% using sample similarity method.
Table~\ref{table:voice} shows several voice-based user authentication methods. The listed voice-based methods show the validity of using this modality for the user authentication task.

\BfPara{Insights and Challenges}
The high availability of voice recognition systems enables a simple and accurate implementation of voice-based authentication schemes.
However, there are many shortcomings when relying solely on the voice-based modality for user authentication. Therefore, many studies have employed voice in multimodal authentication approaches \cite{zhou2019multi,morris2006multimodal,gofman2016multimodal,kim2008multimodal,gofman2016hidden}.
These shortcomings include the following. \cib{1} \textbf{Background Noise:} Voice samples captured by mobile devices usually contain noises, considering the mobility and uncontrolled environmental conditions. \cib{2} \textbf{User Physical and Emotional State:} Changes in the voice caused by emotions or illness (\ie throat-related conditions) may affect the performance of the system. \cib{3} \textbf{Adversarial Attacks:} The rise of adversarial examples suggests the possibility of successfully crafting samples to fool the authentication model and force it to grant access to imposters.  \cib{4} \textbf{System Overhead:} Continuous voice-based user authentication methods require voice commands and signatures to be captured and analyzed periodically through a sophisticated system with multiple stages that include data collection, noise reduction, and voice recognition. Such processes introduce overhead in terms of both power and computation. \cib{5} \textbf{Usability:} Considering the user and environmental changes and the variety of possible noise sources, voice-based methods may result in high false acceptance and false rejection rates. Depending on the sampling rate, the high false rejection rates can degrade the usability and user experience. All of those issues require further attention through additional research efforts.

\begin{table*}[t]
\centering
\caption{Summary of the related work for multimodal-based user authentication. Each work is identified by the used modalities, utilized sensors, dataset, modeling algorithm, and their performance.}
\label{table:multimodal}
\scalebox{0.97}{\begin{tabular}{lllccccccccc}
\toprule
Study & Modalities & Sensors &  Methods & \begin{tabular}[c]{@{}c@{}}\# \\ Users\end{tabular} & EER & FAR & FRR & TPR &  Accuracy & \begin{tabular}[c]{@{}c@{}}Auth. \\ Time\end{tabular} \\
\midrule

\cite{gofman2016hidden}   &   Face/Voice  &      Ca, Mi    & LDA-HMM & 54 & \err{22}21.58\% & \xmark & \xmark & \cmark & \xmark & 0.39s\\

\cite{kim2008multimodal}   &   Teeth Images/voice  &      Ca, Mi    & HMM$\oplus$GMM & 50 & \err{2}2.13\% & \cmark & \cmark & \xmark & \xmark & 10.76s \\

\cite{gofman2016multimodal}   &   Face/Voice  &      NA    & LDA-Matching & 54 & \err{2}2.14\% & \xmark & \xmark & \xmark & \xmark & 1.57s \\

\cite{morris2006multimodal}   &   Face/Voice/Signature  &      NA    & GMM & 60 & \err{1} 0.56\% & 0.97\% & 0.69\% & \xmark & \xmark & \xmark \\

\cite{khamis2016gazetouchpass}   &   Touch/Gaze  &      To, Ca    & \xmark & 13 & \xmark & \xmark & \xmark & \xmark & \cc{95}65\% & 3.1s \\

\cite{saevanee2015continuous}   &   Keystroke/Linguistic/Behavior  &      NA    & MLP$\oplus$RBF & 30 & \err{3}3.3\% & \xmark & \xmark & \xmark & \xmark & 2--10m \\

\cite{neal2015mobile}   &   App/Bluetooth/Wi-Fi  &      NA    & k-NN & 200 & \xmark & \xmark & \xmark & \xmark & \cc{85}85\% & \xmark \\

\cite{stanciu2016effectiveness}   &   Keystroke/Sensor dynamics  &      To, Ac, Gy    & k-NN & 20 & \err{0}0.14\% & \xmark & \xmark & \xmark & \xmark & \xmark \\

\cite{sitova2015hmog}   &   Keystroke/Motion/Orientation  &      To, Ac, Gy    & PCA-SVM & 20 &\err{7} 7.16\% & \cmark & \cmark & \xmark & \xmark & 20s \\

\cite{raja2015multi}   &   Face/Periocular/Iris  &      Ca    & FA-NN & 78 & \err{1}0.68\% & \cmark & \xmark & \xmark & \cmark & \xmark \\

\cite{stokkenes2017feature}   &   Face/Periocular  &      Ca    & Matching & 73 & \err{1}1.34\% &  0.01\% & \xmark & \xmark & \cc{95}94.66\% & \xmark \\

\cite{tiong2017multimodal}   &   Face/Periocular  &      Ca    & CNN & 246 & \xmark &  \xmark & \xmark & \cmark & \cc{99}98.5\% & \xmark \\

\cite{lamiche2019continuous}   &   Keystroke/Gait  &      To, Ac    & MLP & 20 & \err{1}1\% &  0.68\% & 7\% & \xmark & \cc{99}99.11\% & \xmark \\

\cite{pang2019mineauth}   &   App/Bluetooth/Wi-Fi/other  &     NA    & FPOS & 33 & \xmark & \cmark & \xmark & \cmark & \cc{98}98.3\% & 2.3s \\

\cite{acien2019multilock}   &   Touch/Motion/App/other  &     {\scriptsize To, Ac, Gy, Ma, Li}    & SVM & 48 & \cmark & \cmark & \cmark & \xmark & \cc{97}97.1\% & \xmark \\

\cite{KayacikJBAM14}   &   App/Motion/Wi-Fi/other &   {\scriptsize Ac, WiFi, Li, other} & Ensemble & 7 & \xmark & \xmark & \xmark & \xmark & \cc{99}99.4\% & 122s \\

\cite{ZhuWWZ13}   &   Motion/Gesture &   Ac, Gr, Or, Ma & n-gram & 20 & \xmark & 31.1\% & \xmark & 71.30\% & \xmark & 4.96s \\

\cite{FenuM18}   &   Face/Touch/Motion &     {\scriptsize Ca, To, Ac,Gy, Ma}    & Ensemble & 100 & \err{4}0.8-3.6\% & \xmark & \xmark & \xmark & \xmark & \cmark \\

\cite{lin2018tdsd}   &   Touch/Motion/other  &    {\scriptsize To, Ac, Gy, Ma, other}    & {\scriptsize Compound-Voting} & 30 & \xmark & 0 & 0 & \xmark & \cc{100}100\% & \xmark \\

\cite{volaka2019towards}   &   Touch/Motion &     To, Ac, Gy    & SVM & 100 & \err{15}15\% & \xmark & \xmark & \xmark & \cc{88}88\% & \xmark \\

\cite{ShenYYLG16}   &   Touch/Motion &     To, Ac, Gy    & SVM & 48 & \cmark & 5.01\% & 6.85\% & \xmark & \xmark & \cmark \\

\cite{zhou2019multi}   &   Face/Voice &     Ca, Mi    & CNN-SVM & 10 & \xmark & \xmark & \xmark & 88.84\% & \cc{94} 94.07\% & 30ms \\

\cite{centeno2018mobile}   &   Touch/Motion &   Ac, Gy, Ma    & CNN-SVM & 90 & \xmark & \cmark & \cmark & \xmark & \cc{98}97.8\% & 1s \\

\cite{AbuhamadAMN20}   &   Touch/Motion &   Ac, Gy, Ma, El    & LSTM & 84 & \err{0}0.37\% & 1.72\% & 8.47\% & \cmark & \cc{98}97.84\% & 1s \\

\cite{ShenLCGM18}   &   Touch/Motion &   {\scriptsize To, Ac, Gy, Ma, Or}    & HMM & 102 & \err{5}4.74\% & 3.98\% & 5.03\% & \xmark & \xmark & 8s \\

\cite{LeeL17}   &   Wearable/Sensor dynamics &   {\scriptsize Ac, Gy, Ma, Or, Li}    & KRR & 35 & \cmark & 2.8\% & 0.9\% & \cmark & \cc{98}98.1\% & \cmark \\

\cite{ZhuQXZS19}   &   Wearable/Sensor dynamics &   {\scriptsize Ac, Gy, Gr}    & SVM & 1,513 & \xmark & \xmark & \xmark & 73.28\% & \cc{96} 95.57\% & 3.2s \\

\cite{MoseniaSRJ17}   &   Healthcare readings &   HWS    & SVM-RBF & 37 & \err{3}2.6\%& 7.6\% & 9.6\% & \cmark & \xmark & \cmark \\

{}                      &   {} &   {}    & AdaBoost & 37 & \err{2} 2.4\% & 7.6\% & 8.4\% & \cmark & \xmark & \cmark\\

\midrule

\multicolumn{11}{l}{Ca: Camera, Mi: Microphone, To: Touch, Ac: Accelerometer, Gy: Gyroscope, Ma: Magnetometer, Li: Light sensor, Gr: Gravity sensor, El: Elevation}\\
\multicolumn{11}{l}{HWS: Healthcare Wearable Sensors, LDA: Linear Discriminant Analysis, HMM: Hidden Markov Model, GMM: Gaussian Mixed Model,  } \\
\multicolumn{11}{l}{MLP: Multilayer Perceptron, RBF: Radial Basis Function, k-NN: k-Nearest Neighbor, PCA: Principal Component Analysis, SVM: Support Vector Machine,} \\
\multicolumn{11}{l}{FA-NN: Fast Approximate Nearest Neighbor, CNN: Convolutional Neural Network, FPOS: Frequent Pattern Outlier Score, KRR: Kernel Ridge Regression.} \\

\bottomrule
\end{tabular}}
\end{table*}

\section{Multimodal Authentication}\label{sec:multimodalAuthentication}

Multimodal authentication systems have become increasingly popular since relying on multiple modalities on offer robust and accurate results in comparison to unimodal systems, that consider only a single biometric modality. Such systems offer hardened security, especially against adversarial attacks, and deliver a flexible method for authentication considering possible changes of the input data that result in problems in the enrollment and validation phase \cite{jain2004introduction,BaQFR19}. 

The implementation of multimodal authentication could require a fusion of multiple data sources, extracted features, or/and used algorithms and models. The literature shows that multimodal biometric-based authentication schemes have used different fusion approaches such as feature-level fusion, used modeling algorithms fusion, and decision-level fusion. \cib{1} Feature-level fusion includes combining features from different modalities to be considered together as an input to the modeling algorithm. Accounting for possible heterogeneous resulting feature space from different sources, a normalization process usually takes place. \cib{2} Algorithm-level fusion includes constructing an ensemble of models that are built based on an individual of multiple biometric modalities. The ensemble combines outputs by considering matching scores or voting mechanism to help with the decision. \cib{3} Decision-level fusion occurs when decisions are generated by individual modalities separately. The final decision considers all outputs and adopts certain rules or voting to generate the final output.

Using multimodal authentication on smartphones is a feasible solution since today's devices are equipped with a variety of sensors that support the reading of several biometrics \cite{gofman2016multimodal}. However, several challenges should be considered when implementing multimodal authentication, such as the input data quality generated by different sources since poor data results in poor performance, and the inclusion of multiple data sources requires reading from different sensors, which could be computationally-hungry and energy-expensive \cite{jain2006biometrics}. Addressing such challenges effectively allows multimodal authentication to offer robust and secure access control \cite{rahman2014seeing}. 

Vildjiounaite \etal~\cite{vildjiounaite2006unobtrusive} proposed combining gait and voice biometrics to increase the performance of user validation. Using data samples of 31 users, the authors reported a decrease in the error rates from 2.82\%–43.09\% and 13.7–17.2\% using the individual voice and gait recognition, respectively,  to 1.97\%–11.8\% for adopting a multimodal system incorporating both biometrics. 
However, the proposed method is event-dependant and performs differently when the user motion or speaking is different since the results show that such a method is ineffective if the user is not speaking or speaking.
Zhu \etal \cite{ZhuQXZS19} proposed an SVM-based method called {\em RiskCog} that can validate users within 3.2  seconds using sensory data collected from mobile and/or wearable devices including readings of the accelerometer, gyroscope and gravity sensors. The authors reported an average system accuracy of 93.8\% and 95.6\% for steady and moving users, respectively, using a large dataset of 1,513 users. 
Lee \etal \cite{LeeL17} proposed combining sensors' readings from the user's smartphone and other wearable devices to improve authentication accuracy. Their experiments on a dataset of 35 users have shown an accuracy of 98.1\%, FRR of 0.9\%, and FAR of 2.8\% by combining data from users' smartphones and smartwatches when adopting an authentication window of six seconds.

Gofman \etal~\cite{gofman2016multimodal} suggested using face and voice biometrics to tackle input data quality and training data scarcity for mobile authentication. Considering the nature of the data acquisition process on mobile devices, the authors argued that data quality is usually in poor condition due to environmental factors or the utilization of low-cost sensors. Moreover, the authors stated mobile authentication systems face a training data scarcity problem since users tend to provide small training samples during the enrolment phase. Using a multimodal system, the authors addressed these issues and enhanced the potential of acquiring high-quality data samples during user enrollment. The proposed approach incorporated the Fisherface method for face recognition since it is shown to be effective under changing environmental conditions, and Hidden Markov Models (HMM) and Linear Discriminant Analysis (LDA) for voice recognition (HMM was used for algorithm score-level fusion and LDA was used for feature-level fusion ). The authors used a quality-based weighting method to adjust to samples' quality and limit the impact of poor-quality samples on the performance of the system. The results showed a decrease in error rates from 4.29\% for the face recognition module and 34.72\% for the voice recognition module to 2.14\% for the feature-level fused multimodal system.
Similar work has been proposed by Morris \etal~\cite{morris2006multimodal} for combining voice, face, and signature modalities for personal digital assistant devices. The authors reported a decrease in error rates when combining all three modalities from 3.38\%-29.87\% to 0.56\%, which is considered a considerable improvement in the system performance. Their implementation adopts a text-dependent voice authentication approach since text-independent can bring much complexity when addressing phonetic variations, which can computationally-expensive and energy-draining when running locally on the device.

Kayacik \etal \cite{KayacikJBAM14} proposed a data-driven approach with an ensemble of classifiers to enable the authentication system to be temporally and spatially aware of the user behavioral usage and surroundings by taking advantage of several hard and soft sensors such as the accelerometer, WiFi, light sensor, and others. The proposed method requires more than 122 seconds to allow the data to be collected for authenticating users and about 717 seconds to detect an imposter. However, the experiments report a high authentication accuracy of 99.4\%.
Similar work has been proposed by Li and Bours \cite{LiB18} that incorporates sensory data of smartphones and WiFi information for enabling users to access an application within three seconds, with an average EER of 9.19\%. 
Similar studies combinations of multiple biometrics to incorporate face, iris, and periocular recognition \cite{raja2015multi,raja2015fusion},  eye gaze, and touch gestures \cite{khamis2016gazetouchpass}, and user behavioral profiling, keystroke dynamics, and linguistic features \cite{saevanee2015continuous}. Another direction of research studied users' behavioral patterns using their usage of applications and Wi-Fi traffic \cite{neal2015mobile}.
Table~\ref{table:multimodal} shows the multimodal-based user authentication methods by using multiple modalities and machine learning algorithms. 

\BfPara{Insights and Challenges}
Multimodal-based user authentication methods are designed by implementing several modalities that can include both behavioral and physiological biometrics (\eg face, voice, and keystroke dynamics) to conduct user authentication tasks. 
Recent trends in the authentication space show that multimodal methods are the favorable choice for implementing authentication schemes due to their performance and added security.
Since multimodal authentication schemes incorporate multiple modalities, they intrinsically inherit some of the shortcomings and challenges of their integrated components. However, adopting a multimodal authentication scheme for continuous authentication on mobile devices adds several additional challenges, among which we mention the following.
\cib{1} \textbf{Computation and Memory Overhead:} Incorporating multiple modalities requires continuous collection and processing of data at a high sampling rate, which can increase the computation and memory overhead of the device. Moreover, combining the output of multiple modalities for the authentication decision requires the inference of multiple models or matching algorithms to generate the final output. Considering continuous authentication at a high frequency can introduce major resources bottlenecks, in terms of computations. Fortunately, 
current mobile devices are equipped with multi-core processors, GPUs, and even Gigabytes of RAM, making it feasible to run a wide range of sophisticated applications such as multimodal-based continuous authentication schemes. Recent trends to secure in-device operations take advantage of machine learning libraries that utilize hardware acceleration units, using GPUs or Digital Signal Processor (DSPs) which are available in most of today's mobile devices, to implement local inference of authentication models.
\cib{2} \textbf{Biometric Samples Quality Assurance:}
The performance of a system is related to the quality of the collected samples, as a biometric sample with a high quality is essential for an accurate identification. Due to the unreliable features that could be obtained from a single biometric (\ie the changing emotional or physical state of the user or poor data acquisition), and to overcome performance degradation caused by these limitations, researchers have moved from the unimodal to multimodal biometrics. For instance, combining face recognition and keystroke dynamics for user authentication enhances the performance of each modality when considered alone. However, recent trends in adopting biometric-based authentication show it is also necessary to add a sample-quality assessment module to the authentication system, after the data collection and acquisition module, in order to guarantee the processing of valid samples in further processes. 
\cib{3} \textbf{Machine Learning-based Authentication:}
Recent studies show the increasing reliance on machine learning techniques to implement  authentication systems. For multimodal-based methods, researchers utilize an ensemble of machine learning models to enable multiple pattern recognitions per legitimate user. 
This can result in a longer training time (\ie extending the user enrolment phase), greater model size and memory overhead, and inference time (\ie user authentication phase). All of those are open directions worth exploring. Especially, future authentication schemes should consider using hardware acceleration units, such as GPUs or DSPs that are available in most of today's mobile devices.

\section{Conclusion}\label{sec:conclusion}
Mobile devices have become the most common platform for communication and accessing the internet. The rapid enhancements of embedded technologies and resources of mobile devices have enabled users to conduct varieties of activities and transactions. Therefore, secure and accurate access control is essential. To date, mobile devices' manufacturers have implemented knowledge-based and physiological biometric-based authentication methods as the primary access control scheme. While both approaches offer simplicity, efficiency, and precision, they assume the same level of security to all applications and fall short on delivering authentication beyond the point-of-entry. Moreover, these approaches require overt recognition, where the user explicitly enters the pass-secret or the used biometrics, making them fail in delivering implicit, transparent, and continuous authentication. Recently, behavioral biometrics are used to offer efficient continuous authentication on smartphones by leveraging the readings of a variety of embedded sensors. This survey aims to highlights methods, approaches, benefits, and challenges associated with using behavioral biometrics for user authentication. We surveyed around 150 studies that conducted a behavioral-based authentication and pointed out their used techniques, sensors, performance measurements, and time needed for authentication. As this field is rapidly evolving, there is still a need to explore security-related aspects and implementation considerations beyond familiar standards.

\balance

\bibliographystyle{IEEEtran}
\bibliography{ref}

\begin{thebibliography}{100}
\providecommand{\url}[1]{#1}
\csname url@samestyle\endcsname
\providecommand{\newblock}{\relax}
\providecommand{\bibinfo}[2]{#2}
\providecommand{\BIBentrySTDinterwordspacing}{\spaceskip=0pt\relax}
\providecommand{\BIBentryALTinterwordstretchfactor}{4}
\providecommand{\BIBentryALTinterwordspacing}{\spaceskip=\fontdimen2\font plus
\BIBentryALTinterwordstretchfactor\fontdimen3\font minus
  \fontdimen4\font\relax}
\providecommand{\BIBforeignlanguage}[2]{{%
\expandafter\ifx\csname l@#1\endcsname\relax
\typeout{** WARNING: IEEEtran.bst: No hyphenation pattern has been}%
\typeout{** loaded for the language `#1'. Using the pattern for}%
\typeout{** the default language instead.}%
\else
\language=\csname l@#1\endcsname
\fi
#2}}
\providecommand{\BIBdecl}{\relax}
\BIBdecl

\bibitem{jung2018digitalseal}
C.~Jung, J.~Kang, A.~Mohaisen, and D.~Nyang, ``Digitalseal: a transaction
  authentication tool for online and offline transactions,'' in \emph{2018 IEEE
  International Conference on Acoustics, Speech and Signal Processing
  (ICASSP)}.\hskip 1em plus 0.5em minus 0.4em\relax IEEE, 2018, pp. 6956--6960.

\bibitem{AlAbdulwahidCSFR16}
A.~Al~Abdulwahid, N.~Clarke, I.~Stengel, S.~Furnell, and C.~Reich, ``Continuous
  and transparent multimodal authentication: Reviewing the state of the art,''
  \emph{Cluster Computing}, vol.~19, no.~1, Mar. 2016.

\bibitem{neal2016surveying}
T.~J. Neal and D.~L. Woodard, ``Surveying biometric authentication for mobile
  device security,'' \emph{Journal of Pattern Recognition Research}, vol.~1,
  pp. 74--110, 2016.

\bibitem{zhao2015picture}
Z.~Zhao, G.-J. Ahn, and H.~Hu, ``Picture gesture authentication: Empirical
  analysis, automated attacks, and scheme evaluation,'' \emph{ACM Transactions
  on Information and System Security (TISSEC)}, vol.~17, no.~4, p.~14, 2015.

\bibitem{nyang2014keylogging}
D.~Nyang, A.~Mohaisen, and J.~Kang, ``Keylogging-resistant visual
  authentication protocols,'' \emph{IEEE Transactions on Mobile Computing},
  vol.~13, no.~11, pp. 2566--2579, 2014.

\bibitem{nyang2018two}
D.~Nyang, H.~Kim, W.~Lee, S.-b. Kang, G.~Cho, M.-K. Lee, and A.~Mohaisen,
  ``Two-thumbs-up: Physical protection for pin entry secure against recording
  attacks,'' \emph{computers \& security}, vol.~78, pp. 1--15, 2018.

\bibitem{de2012touch}
A.~De~Luca, A.~Hang, F.~Brudy, C.~Lindner, and H.~Hussmann, ``Touch me once and
  i know it's you!: implicit authentication based on touch screen patterns,''
  in \emph{proceedings of the SIGCHI Conference on Human Factors in Computing
  Systems}.\hskip 1em plus 0.5em minus 0.4em\relax ACM, 2012, pp. 987--996.

\bibitem{clarke2005authentication}
N.~L. Clarke and S.~M. Furnell, ``Authentication of users on mobile
  telephones--a survey of attitudes and practices,'' \emph{Computers \&
  Security}, vol.~24, no.~7, pp. 519--527, 2005.

\bibitem{amin2014biometric}
R.~Amin, T.~Gaber, G.~ElTaweel, and A.~E. Hassanien, ``Biometric and
  traditional mobile authentication techniques: Overviews and open issues,'' in
  \emph{Bio-inspiring cyber security and cloud services: trends and
  innovations}.\hskip 1em plus 0.5em minus 0.4em\relax Springer, 2014, pp.
  423--446.

\bibitem{crawford2014understanding}
H.~Crawford and K.~Renaud, ``Understanding user perceptions of transparent
  authentication on a mobile device,'' \emph{Journal of Trust Management},
  vol.~1, no.~1, p.~7, 2014.

\bibitem{furnell2008beyond}
S.~Furnell, N.~Clarke, and S.~Karatzouni, ``Beyond the pin: Enhancing user
  authentication for mobile devices,'' \emph{Computer fraud \& security}, vol.
  2008, no.~8, pp. 12--17, 2008.

\bibitem{khan2014towards}
H.~Khan and U.~Hengartner, ``Towards application-centric implicit
  authentication on smartphones,'' in \emph{Proceedings of the 15th Workshop on
  Mobile Computing Systems and Applications}.\hskip 1em plus 0.5em minus
  0.4em\relax ACM, 2014, p.~10.

\bibitem{wojtowicz2016model}
A.~W{\'o}jtowicz and K.~Joachimiak, ``Model for adaptable context-based
  biometric authentication for mobile devices,'' \emph{Personal and Ubiquitous
  Computing}, vol.~20, no.~2, pp. 195--207, 2016.

\bibitem{yu2016usable}
Z.~Yu, I.~Olade, H.-N. Liang, and C.~Fleming, ``Usable authentication
  mechanisms for mobile devices: An exploration of 3d graphical passwords,'' in
  \emph{2016 International Conference on Platform Technology and Service
  (PlatCon)}.\hskip 1em plus 0.5em minus 0.4em\relax IEEE, 2016, pp. 1--3.

\bibitem{shin2012design}
K.~I. Shin, J.~S. Park, J.~Y. Lee, and J.~H. Park, ``Design and implementation
  of improved authentication system for android smartphone users,'' in
  \emph{2012 26th International Conference on Advanced Information Networking
  and Applications Workshops}.\hskip 1em plus 0.5em minus 0.4em\relax IEEE,
  2012, pp. 704--707.

\bibitem{yang2016free}
Y.~Yang, G.~D. Clark, J.~Lindqvist, and A.~Oulasvirta, ``Free-form gesture
  authentication in the wild,'' in \emph{Proceedings of the 2016 CHI Conference
  on Human Factors in Computing Systems}.\hskip 1em plus 0.5em minus
  0.4em\relax ACM, 2016, pp. 3722--3735.

\bibitem{li2014active}
F.~Li, N.~Clarke, M.~Papadaki, and P.~Dowland, ``Active authentication for
  mobile devices utilising behaviour profiling,'' \emph{International journal
  of information security}, vol.~13, no.~3, pp. 229--244, 2014.

\bibitem{zhou2019multi}
B.~Zhou, Z.~Xie, and F.~Ye, ``Multi-modal face authentication using deep visual
  and acoustic features,'' in \emph{ICC 2019-2019 IEEE International Conference
  on Communications (ICC)}.\hskip 1em plus 0.5em minus 0.4em\relax IEEE, 2019,
  pp. 1--6.

\bibitem{cilia2018multi}
D.~Cilia and F.~Inguanez, ``Multi-model authentication using keystroke dynamics
  for smartphones,'' in \emph{2018 IEEE 8th International Conference on
  Consumer Electronics-Berlin (ICCE-Berlin)}.\hskip 1em plus 0.5em minus
  0.4em\relax IEEE, 2018, pp. 1--6.

\bibitem{miles2006tracking}
C.~A. Miles and J.~P. Cohn, ``Tracking prisoners in jail with biometrics: An
  experiment in a navy brig,'' \emph{National Institute of Justice Journal},
  vol. 253, 2006.

\bibitem{schaffer2015expanding}
K.~B. Schaffer, ``Expanding continuous authentication with mobile devices,''
  \emph{Computer}, vol.~48, no.~11, pp. 92--95, 2015.

\bibitem{teh2013survey}
P.~S. Teh, A.~B.~J. Teoh, and S.~Yue, ``A survey of keystroke dynamics
  biometrics,'' \emph{The Scientific World Journal}, vol. 2013, 2013.

\bibitem{6496441}
S.~{Bhatt} and T.~{Santhanam}, ``Keystroke dynamics for biometric
  authentication — a survey,'' in \emph{2013 International Conference on
  Pattern Recognition, Informatics and Mobile Engineering}, Feb 2013, pp.
  17--23.

\bibitem{karnan2011biometric}
M.~Karnan, M.~Akila, and N.~Krishnaraj, ``Biometric personal authentication
  using keystroke dynamics: A review,'' \emph{Applied soft computing}, vol.~11,
  no.~2, pp. 1565--1573, 2011.

\bibitem{tirumala2017speaker}
S.~S. Tirumala, S.~R. Shahamiri, A.~S. Garhwal, and R.~Wang, ``Speaker
  identification features extraction methods: A systematic review,''
  \emph{Expert Systems with Applications}, vol.~90, pp. 250--271, 2017.

\bibitem{Spolaor2016}
R.~Spolaor, L.~QianQian, M.~Monaro, M.~Conti, L.~Gamberini, and G.~Sartori,
  ``Biometric authentication methods on smartphones: A survey.''
  \emph{PsychNology Journal}, vol.~14, no.~2, pp. 87 -- 98, 2016.

\bibitem{kunda2018survey}
D.~Kunda and M.~Chishimba, ``A survey of android mobile phone authentication
  schemes,'' \emph{Mobile Networks and Applications}, pp. 1--9, 2018.

\bibitem{EhatishamANAL18}
M.~Ehatisham-ul Haq, M.~Awais~Azam, U.~Naeem, Y.~Amin, and J.~Loo, ``Continuous
  authentication of smartphone users based on activity pattern recognition
  using passive mobile sensing,'' \emph{J. Netw. Comput. Appl.}, vol. 109,
  no.~C, May 2018.

\bibitem{NwekeTAA18}
H.~F. Nweke, Y.~W. Teh, M.~A. Al-Garadi, and U.~R. Alo, ``Deep learning
  algorithms for human activity recognition using mobile and wearable sensor
  networks: State of the art and research challenges,'' \emph{Expert Systems
  with Applications}, 2018.

\bibitem{AmanBS19}
M.~N. Aman, M.~H. Basheer, and B.~Sikdar, ``Two-factor authentication for iot
  with location information,'' \emph{{IEEE} Internet of Things Journal},
  vol.~6, no.~2, pp. 3335--3351, 2019.

\bibitem{ShenLCGM18}
C.~Shen, Y.~Li, Y.~Chen, X.~Guan, and R.~A. Maxion, ``Performance analysis of
  multi-motion sensor behavior for active smartphone authentication,''
  \emph{{IEEE} Trans. Information Forensics and Security}, vol.~13, no.~1, pp.
  48--62, 2018.

\bibitem{ZhangHXCH17}
Y.~Zhang, W.~Hu, W.~Xu, C.~T. Chou, and J.~Hu, ``Continuous authentication
  using eye movement response of implicit visual stimuli,'' \emph{Proc. ACM
  Interact. Mob. Wearable Ubiquitous Technol., {IMWUT}}, vol.~1, no.~4, Jan.
  2018.

\bibitem{Arteaga-Falconi16}
J.~S. Arteaga{-}Falconi, H.~A. Osman, and A.~El{-}Saddik, ``{ECG}
  authentication for mobile devices,'' \emph{{IEEE} Trans. Instrumentation and
  Measurement}, vol.~65, no.~3, pp. 591--600, 2016.

\bibitem{BaPFKMR18}
Z.~Ba, S.~Piao, X.~Fu, D.~Koutsonikolas, A.~Mohaisen, and K.~Ren, ``Abc:
  enabling smartphone authentication with built-in camera,'' in \emph{25th
  Annual Network and Distributed System Security Symposium, NDSS 2018}, 2018.

\bibitem{ZengNYMZWZ14}
M.~Zeng, L.~T. Nguyen, B.~Yu, O.~J. Mengshoel, J.~Zhu, P.~Wu, and J.~Zhang,
  ``Convolutional neural networks for human activity recognition using mobile
  sensors,'' in \emph{6th International Conference on Mobile Computing,
  Applications and Services, MobiCASE}, 2014, pp. 197--205.

\bibitem{BiegelC04}
G.~Biegel and V.~Cahill, ``A framework for developing mobile, context-aware
  applications,'' in \emph{Proceedings of the Second {IEEE} International
  Conference on Pervasive Computing and Communications (PerCom)}, 2004, pp.
  361--365.

\bibitem{DasGD16}
S.~Das, D.~Guha, and B.~Dutta, ``Medical diagnosis with the aid of using fuzzy
  logic and intuitionistic fuzzy logic,'' \emph{Applied Intelligence}, vol.~45,
  no.~3, pp. 850--867, 2016.

\bibitem{ChoiKCPKWK17}
S.~Choi, D.~J. Kim, Y.~Y. Choi, K.~Park, S.-W. Kim, S.~H. Woo, and J.~J. Kim,
  ``A multisensor mobile interface for industrial environment and healthcare
  monitoring,'' \emph{IEEE Transactions on Industrial Electronics}, vol.~64,
  no.~3, pp. 2344--2352, 2017.

\bibitem{WuJ19}
J.~{Wu} and R.~{Jafari}, ``Orientation independent activity/gesture recognition
  using wearable motion sensors,'' \emph{IEEE Internet of Things Journal},
  vol.~6, no.~2, pp. 1427--1437, April 2019.

\bibitem{YuXDGY19}
Z.~{Yu}, E.~{Xu}, H.~{Du}, B.~{Guo}, and L.~{Yao}, ``Inferring user profile
  attributes from multidimensional mobile phone sensory data,'' \emph{IEEE
  Internet of Things Journal}, vol.~6, no.~3, pp. 5152--5162, June 2019.

\bibitem{micallef2015sensor}
N.~Micallef, H.~G. Kayac{\i}k, M.~Just, L.~Baillie, and D.~Aspinall, ``Sensor
  use and usefulness: Trade-offs for data-driven authentication on mobile
  devices,'' in \emph{2015 IEEE International Conference on Pervasive Computing
  and Communications (PerCom)}.\hskip 1em plus 0.5em minus 0.4em\relax IEEE,
  2015, pp. 189--197.

\bibitem{drosou2012activity}
A.~Drosou and D.~Tzovaras, ``Activity and event related biometrics,'' in
  \emph{Second Generation Biometrics: The Ethical, Legal and Social
  Context}.\hskip 1em plus 0.5em minus 0.4em\relax Springer, 2012, pp.
  129--148.

\bibitem{DraffinZZ13}
B.~Draffin, J.~Zhu, and J.~Y. Zhang, ``Keysens: Passive user authentication
  through micro-behavior modeling of soft keyboard interaction,'' in
  \emph{Mobile Computing, Applications, and Services - 5th International
  Conference, MobiCASE}, 2013, pp. 184--201.

\bibitem{laghari2016biometric}
A.~Laghari, Z.~A. Memon \emph{et~al.}, ``Biometric authentication technique
  using smartphone sensor,'' in \emph{2016 13th International Bhurban
  Conference on Applied Sciences and Technology (IBCAST)}.\hskip 1em plus 0.5em
  minus 0.4em\relax IEEE, 2016, pp. 381--384.

\bibitem{ShenDXLWCW20}
Y.~Shen, B.~Du, W.~Xu, C.~Luo, B.~Wei, L.~Cui, and H.~Wen, ``Securing
  cyber-physical social interactions on wrist-worn devices,'' \emph{ACM
  Transactions on Sensor Networks (TOSN)}, vol.~16, no.~2, pp. 1--22, 2020.

\bibitem{ShenYDXLW18}
Y.~Shen, F.~Yang, B.~Du, W.~Xu, C.~Luo, and H.~Wen, ``Shake-n-shack: Enabling
  secure data exchange between smart wearables via handshakes,'' in \emph{2018
  IEEE International Conference on Pervasive Computing and Communications
  (PerCom)}, 2018, pp. 1--10.

\bibitem{AbuhamadAMN20}
M.~Abuhamad, T.~Abuhmed, D.~Mohaisen, and D.~H. Nyang, ``Autosen: Deep
  learning-based implicit continuous authentication using smartphone sensors,''
  \emph{IEEE Internet of Things Journal}, 2020.

\bibitem{LopezLSS16}
P.~Fernandez-Lopez, J.~Liu-Jimenez, C.~Sanchez-Redondo, and R.~Sanchez-Reillo,
  ``Gait recognition using smartphone,'' in \emph{IEEE International Carnahan
  Conference on Security Technology (ICCST)}, 2016, pp. 1--7.

\bibitem{PozoASC12}
G.~B. Del~Pozo, C.~Sanchez-Avila, A.~De-Santos-Sierra, and J.~Guerra-Casanova,
  ``Speed-independent gait identification for mobile devices,''
  \emph{International Journal of Pattern Recognition and Artificial
  Intelligence}, vol.~26, no.~08, p. 1260013, 2012.

\bibitem{DamaseviciusMVW16}
R.~Dama{\v{s}}evi{\v{c}}ius, R.~Maskeli{\=u}nas, A.~Ven{\v{c}}kauskas, and
  M.~Wo{\'z}niak, ``Smartphone user identity verification using gait
  characteristics,'' \emph{symmetry}, vol.~8, no.~10, p. 100, 2016.

\bibitem{Lu2017}
Y.~Lu, Y.~Wei, L.~Liu, J.~Zhong, L.~Sun, and Y.~Liu, ``Towards unsupervised
  physical activity recognition using smartphone accelerometers,''
  \emph{Multimedia Tools Appl.}, vol.~76, no.~8, Apr. 2017.

\bibitem{NickelBB11}
C.~Nickel, H.~Brandt, and C.~Busch, ``Classification of acceleration data for
  biometric gait recognition on mobile devices,'' \emph{Proceedings of the
  Biometrics Special Interest Group, (BIOSIG)}, 2011.

\bibitem{zaliva2015passive}
V.~Zaliva, W.~Melicher, S.~Saha, and J.~Zhang, ``Passive user identification
  using sequential analysis of proximity information in touchscreen usage
  patterns,'' in \emph{2015 Eighth International Conference on Mobile Computing
  and Ubiquitous Networking (ICMU)}.\hskip 1em plus 0.5em minus 0.4em\relax
  IEEE, 2015, pp. 161--166.

\bibitem{AvivGMBS10}
A.~J. Aviv, K.~L. Gibson, E.~Mossop, M.~Blaze, and J.~M. Smith, ``Smudge
  attacks on smartphone touch screens,'' in \emph{4th {USENIX} Workshop on
  Offensive Technologies, {WOOT}}, 2010.

\bibitem{WuC15}
J.~Wu and Z.~Chen, ``An implicit identity authentication system considering
  changes of gesture based on keystroke behaviors,'' \emph{{IJDSN}}, vol.~11,
  pp. 470\,274:1--470\,274:16, 2015.

\bibitem{BuchouxC08}
A.~Buchoux and N.~L. Clarke, ``Deployment of keystroke analysis on a
  smartphone,'' in \emph{Australian Information Security Management
  Conference}, 2008, p.~48.

\bibitem{LuDWWMMS18}
C.~X. Lu, B.~Du, H.~Wen, S.~Wang, A.~Markham, I.~Martinovic, Y.~Shen, and
  N.~Trigoni, ``Snoopy: Sniffing your smartwatch passwords via deep sequence
  learning,'' \emph{Proceedings of the ACM on Interactive, Mobile, Wearable and
  Ubiquitous Technologies}, vol.~1, no.~4, pp. 1--29, 2018.

\bibitem{Mondal2017}
S.~Mondal and P.~Bours, ``Person identification by keystroke dynamics using
  pairwise user coupling,'' \emph{IEEE Transactions on Information Forensics
  and Security}, vol.~12, no.~6, pp. 1319--1329, June 2017.

\bibitem{KambourakisDPP16}
G.~Kambourakis, D.~Damopoulos, D.~Papamartzivanos, and E.~Pavlidakis,
  ``Introducing touchstroke: keystroke-based authentication system for
  smartphones,'' \emph{Security and Communication Networks}, vol.~9, no.~6, pp.
  542--554, 2016.

\bibitem{ReynoldsQD00}
D.~A. Reynolds, T.~F. Quatieri, and R.~B. Dunn, ``Speaker verification using
  adapted gaussian mixture models,'' \emph{Digital signal processing}, vol.~10,
  no. 1-3, pp. 19--41, 2000.

\bibitem{LuBPKL11}
H.~Lu, A.~J.~B. Brush, B.~Priyantha, A.~K. Karlson, and J.~Liu, ``Speakersense:
  Energy efficient unobtrusive speaker identification on mobile phones,'' in
  \emph{Proceedings of the 9th International Conference on Pervasive
  Computing}, 2011, pp. 188--205.

\bibitem{gofman2016hidden}
M.~I. Gofman, S.~Mitra, and N.~Smith, ``Hidden markov models for feature-level
  fusion of biometrics on mobile devices,'' in \emph{2016 IEEE/ACS 13th
  International Conference of Computer Systems and Applications
  (AICCSA)}.\hskip 1em plus 0.5em minus 0.4em\relax IEEE, 2016, pp. 1--2.

\bibitem{ClarkeM07}
N.~L. Clarke and A.~Mekala, ``The application of signature recognition to
  transparent handwriting verification for mobile devices,'' \emph{Information
  management \& computer security}, vol.~15, no.~3, pp. 214--225, 2007.

\bibitem{Martinez-DiazFGO08}
M.~Martinez-Diaz, J.~Fierrez, J.~Galbally, and J.~Ortega-Garcia, ``Towards
  mobile authentication using dynamic signature verification: useful features
  and performance evaluation,'' in \emph{19th International Conference on
  Pattern Recognition}, 2008, pp. 1--5.

\bibitem{morris2006multimodal}
A.~C. Morris, S.~Jassim, H.~Sellahewa, L.~Allano, J.~Ehlers, D.~Wu, J.~Koreman,
  S.~Garcia-Salicetti, B.~Ly-Van, and B.~Dorizzi, ``Multimodal person
  authentication on a smartphone under realistic conditions,'' in \emph{Mobile
  Multimedia/Image Processing for Military and Security Applications}, vol.
  6250.\hskip 1em plus 0.5em minus 0.4em\relax International Society for Optics
  and Photonics, 2006, p. 62500D.

\bibitem{AlzubaidiK16}
A.~Alzubaidi and J.~Kalita, ``Authentication of smartphone users using
  behavioral biometrics,'' \emph{{IEEE} Communications Surveys and Tutorials},
  vol.~18, no.~3, pp. 1998--2026, 2016.

\bibitem{AminiNPGYK18}
S.~Amini, V.~Noroozi, A.~Pande, S.~Gupte, P.~S. Yu, and C.~Kanich, ``Deepauth:
  A framework for continuous user re-authentication in mobile apps,'' in
  \emph{Proceedings of the 27th ACM International Conference on Information and
  Knowledge Management, {CIKM}}, 2018, pp. 2027--2035.

\bibitem{LeeL17}
W.-H. Lee and R.~B. Lee, ``Implicit smartphone user authentication with sensors
  and contextual machine learning,'' in \emph{47th Annual IEEE/IFIP
  International Conference on Dependable Systems and Networks, (DSN)}, 2017,
  pp. 297--308.

\bibitem{crawford2013framework}
H.~Crawford, K.~Renaud, and T.~Storer, ``A framework for continuous,
  transparent mobile device authentication,'' \emph{Computers \& Security},
  vol.~39, pp. 127--136, 2013.

\bibitem{hong2015waving}
F.~Hong, M.~Wei, S.~You, Y.~Feng, and Z.~Guo, ``Waving authentication: your
  smartphone authenticate you on motion gesture,'' in \emph{Proceedings of the
  33rd Annual ACM Conference Extended Abstracts on Human Factors in Computing
  Systems}.\hskip 1em plus 0.5em minus 0.4em\relax ACM, 2015, pp. 263--266.

\bibitem{feng2013investigating}
T.~Feng, X.~Zhao, and W.~Shi, ``Investigating mobile device picking-up motion
  as a novel biometric modality,'' in \emph{2013 IEEE Sixth International
  Conference on Biometrics: Theory, Applications and Systems (BTAS)}.\hskip 1em
  plus 0.5em minus 0.4em\relax IEEE, 2013, pp. 1--6.

\bibitem{wu2015implicit}
J.~Wu and Z.~Chen, ``An implicit identity authentication system considering
  changes of gesture based on keystroke behaviors,'' \emph{International
  Journal of Distributed Sensor Networks}, vol.~11, no.~6, p. 470274, 2015.

\bibitem{hong2016mgra}
F.~Hong, S.~You, M.~Wei, Y.~Zhang, and Z.~Guo, ``Mgra: Motion gesture
  recognition via accelerometer,'' \emph{Sensors}, vol.~16, no.~4, p. 530,
  2016.

\bibitem{lu2018multifactor}
D.~Lu, D.~Huang, Y.~Deng, and A.~Alshamrani, ``Multifactor user authentication
  with in-air-handwriting and hand geometry,'' in \emph{2018 International
  Conference on Biometrics (ICB)}.\hskip 1em plus 0.5em minus 0.4em\relax IEEE,
  2018, pp. 255--262.

\bibitem{xia2018motionhacker}
Q.~Xia, F.~Hong, Y.~Feng, and Z.~Guo, ``Motionhacker: Motion sensor based
  eavesdropping on handwriting via smartwatch,'' in \emph{IEEE INFOCOM
  2018-IEEE Conference on Computer Communications Workshops (INFOCOM
  WKSHPS)}.\hskip 1em plus 0.5em minus 0.4em\relax IEEE, 2018, pp. 468--473.

\bibitem{yan2018towards}
J.~Yan, Y.~Qi, Q.~Rao, and S.~Qi, ``Towards a user-friendly and secure hand
  shaking authentication for smartphones,'' in \emph{2018 17th IEEE
  International Conference On Trust, Security And Privacy In Computing And
  Communications/12th IEEE International Conference On Big Data Science And
  Engineering (TrustCom/BigDataSE)}.\hskip 1em plus 0.5em minus 0.4em\relax
  IEEE, 2018, pp. 1170--1179.

\bibitem{fantana2015movement}
A.~L. Fantana, S.~Ramachandran, C.~H. Schunck, and M.~Talamo, ``Movement based
  biometric authentication with smartphones,'' in \emph{2015 International
  Carnahan Conference on Security Technology (ICCST)}.\hskip 1em plus 0.5em
  minus 0.4em\relax IEEE, 2015, pp. 235--239.

\bibitem{casanova2010real}
J.~G. Casanova, C.~S. {\'A}vila, A.~de~Santos~Sierra, G.~B. del Pozo, and V.~J.
  Vera, ``A real-time in-air signature biometric technique using a mobile
  device embedding an accelerometer,'' in \emph{International Conference on
  Networked Digital Technologies}.\hskip 1em plus 0.5em minus 0.4em\relax
  Springer, 2010, pp. 497--503.

\bibitem{haring2018pick}
M.~Haring, D.~Reinhardt, and Y.~Omlor, ``Pick me up and i will tell you who you
  are: Analyzing pick-up motions to authenticate users,'' in \emph{2018 IEEE
  International Conference on Pervasive Computing and Communications Workshops
  (PerCom Workshops)}.\hskip 1em plus 0.5em minus 0.4em\relax IEEE, 2018, pp.
  472--475.

\bibitem{maghsoudi2016behavioral}
J.~Maghsoudi and C.~C. Tappert, ``A behavioral biometrics user authentication
  study using motion data from android smartphones,'' in \emph{2016 European
  Intelligence and Security Informatics Conference (EISIC)}.\hskip 1em plus
  0.5em minus 0.4em\relax IEEE, 2016, pp. 184--187.

\bibitem{eremin2018touch}
A.~Eremin, K.~Kogos, and Y.~Valatskayte, ``Touch and move: Incoming call user
  authentication,'' in \emph{International Conference on Information Systems
  Security and Privacy}.\hskip 1em plus 0.5em minus 0.4em\relax Springer, 2018,
  pp. 26--39.

\bibitem{LeeL15}
W.~Lee and R.~B. Lee, ``Multi-sensor authentication to improve smartphone
  security,'' in \emph{Proceedings of the 1st International Conference on
  Information Systems Security and Privacy, {ICISSP}}, 2015, pp. 270--280.

\bibitem{LiHZ19}
Y.~{Li}, H.~{Hu}, and G.~{Zhou}, ``Using data augmentation in continuous
  authentication on smartphones,'' \emph{IEEE Internet of Things Journal},
  vol.~6, no.~1, pp. 628--640, Feb 2019.

\bibitem{SongWRX16}
C.~Song, A.~Wang, K.~Ren, and W.~Xu, ``Eyeveri: {A} secure and usable approach
  for smartphone user authentication,'' in \emph{35th Annual {IEEE}
  International Conference on Computer Communications, {INFOCOM}}, 2016, pp.
  1--9.

\bibitem{clarke2007advanced}
N.~L. Clarke and S.~Furnell, ``Advanced user authentication for mobile
  devices,'' \emph{computers \& security}, vol.~26, no.~2, pp. 109--119, 2007.

\bibitem{ferrero2015gait}
R.~Ferrero, F.~Gandino, B.~Montrucchio, M.~Rebaudengo, A.~Velasco, and
  I.~Benkhelifa, ``On gait recognition with smartphone accelerometer,'' in
  \emph{2015 4th Mediterranean Conference on Embedded Computing (MECO)}.\hskip
  1em plus 0.5em minus 0.4em\relax IEEE, 2015, pp. 368--373.

\bibitem{jain2004introduction}
A.~K. Jain, A.~Ross, S.~Prabhakar \emph{et~al.}, ``An introduction to biometric
  recognition,'' \emph{IEEE Transactions on circuits and systems for video
  technology}, vol.~14, no.~1, 2004.

\bibitem{kartik2008multimodal}
P.~Kartik, S.~M. Prasanna, and R.~V. Prasad, ``Multimodal biometric person
  authentication system using speech and signature features,'' in \emph{TENCON
  2008-2008 IEEE Region 10 Conference}.\hskip 1em plus 0.5em minus 0.4em\relax
  IEEE, 2008, pp. 1--6.

\bibitem{bhattacharya2013offline}
I.~Bhattacharya, P.~Ghosh, and S.~Biswas, ``Offline signature verification
  using pixel matching technique,'' \emph{Procedia Technology}, vol.~10, pp.
  970--977, 2013.

\bibitem{sahami2012assessing}
A.~Sahami~Shirazi, P.~Moghadam, H.~Ketabdar, and A.~Schmidt, ``Assessing the
  vulnerability of magnetic gestural authentication to video-based shoulder
  surfing attacks,'' in \emph{Proceedings of the SIGCHI Conference on Human
  Factors in Computing Systems}.\hskip 1em plus 0.5em minus 0.4em\relax ACM,
  2012, pp. 2045--2048.

\bibitem{ZhuHCL17}
H.~Zhu, J.~Hu, S.~Chang, and L.~Lu, ``Shakein: Secure user authentication of
  smartphones with single-handed shakes,'' \emph{IEEE transactions on mobile
  computing}, vol.~16, no.~10, pp. 2901--2912, 2017.

\bibitem{wang2004fusion}
L.~Wang, H.~Ning, T.~Tan, and W.~Hu, ``Fusion of static and dynamic body
  biometrics for gait recognition,'' \emph{IEEE Transactions on circuits and
  systems for video technology}, vol.~14, no.~2, pp. 149--158, 2004.

\bibitem{gafurov2009gait}
D.~Gafurov and E.~Snekkenes, ``Gait recognition using wearable motion recording
  sensors,'' \emph{EURASIP Journal on Advances in Signal Processing}, vol.
  2009, p.~7, 2009.

\bibitem{qian2008people}
G.~Qian, J.~Zhang, and A.~Kidan{\'e}, ``People identification using gait via
  floor pressure sensing and analysis,'' in \emph{European Conference on Smart
  Sensing and Context}.\hskip 1em plus 0.5em minus 0.4em\relax Springer, 2008,
  pp. 83--98.

\bibitem{thang2012gait}
H.~M. Thang, V.~Q. Viet, N.~D. Thuc, and D.~Choi, ``Gait identification using
  accelerometer on mobile phone,'' in \emph{2012 International Conference on
  Control, Automation and Information Sciences (ICCAIS)}.\hskip 1em plus 0.5em
  minus 0.4em\relax IEEE, 2012, pp. 344--348.

\bibitem{hoang2013adaptive}
T.~Hoang, T.~D. Nguyen, C.~Luong, S.~Do, and D.~Choi, ``Adaptive cross-device
  gait recognition using a mobile accelerometer.'' \emph{JIPS}, vol.~9, no.~2,
  p. 333, 2013.

\bibitem{mantyjarvi2005identifying}
J.~Mantyjarvi, M.~Lindholm, E.~Vildjiounaite, S.-M. Makela, and H.~Ailisto,
  ``Identifying users of portable devices from gait pattern with
  accelerometers,'' in \emph{Proceedings.(ICASSP'05). IEEE International
  Conference on Acoustics, Speech, and Signal Processing, 2005.}, vol.~2.\hskip
  1em plus 0.5em minus 0.4em\relax IEEE, 2005, pp. ii--973.

\bibitem{muaaz2013analysis}
M.~Muaaz and R.~Mayrhofer, ``An analysis of different approaches to gait
  recognition using cell phone based accelerometers,'' in \emph{Proceedings of
  International Conference on Advances in Mobile Computing \&
  Multimedia}.\hskip 1em plus 0.5em minus 0.4em\relax ACM, 2013, p. 293.

\bibitem{nickel2011scenario}
C.~Nickel, M.~O. Derawi, P.~Bours, and C.~Busch, ``Scenario test of
  accelerometer-based biometric gait recognition,'' in \emph{2011 Third
  International Workshop on Security and Communication Networks (IWSCN)}.\hskip
  1em plus 0.5em minus 0.4em\relax IEEE, 2011, pp. 15--21.

\bibitem{nickel2012authentication}
C.~Nickel, T.~Wirtl, and C.~Busch, ``Authentication of smartphone users based
  on the way they walk using k-nn algorithm,'' in \emph{2012 Eighth
  International Conference on Intelligent Information Hiding and Multimedia
  Signal Processing}.\hskip 1em plus 0.5em minus 0.4em\relax IEEE, 2012, pp.
  16--20.

\bibitem{hoang2015gait}
T.~Hoang, D.~Choi, and T.~Nguyen, ``Gait authentication on mobile phone using
  biometric cryptosystem and fuzzy commitment scheme,'' \emph{International
  Journal of Information Security}, vol.~14, no.~6, pp. 549--560, 2015.

\bibitem{vildjiounaite2006unobtrusive}
E.~Vildjiounaite, S.-M. M{\"a}kel{\"a}, M.~Lindholm, R.~Riihim{\"a}ki,
  V.~Kyll{\"o}nen, J.~M{\"a}ntyj{\"a}rvi, and H.~Ailisto, ``Unobtrusive
  multimodal biometrics for ensuring privacy and information security with
  personal devices,'' in \emph{International Conference on Pervasive
  Computing}.\hskip 1em plus 0.5em minus 0.4em\relax Springer, 2006, pp.
  187--201.

\bibitem{nickel2013classifying}
C.~Nickel and C.~Busch, ``Classifying accelerometer data via hidden markov
  models to authenticate people by the way they walk,'' \emph{IEEE Aerospace
  and Electronic Systems Magazine}, vol.~28, no.~10, pp. 29--35, 2013.

\bibitem{xu2018keh}
W.~Xu, G.~Lan, Q.~Lin, S.~Khalifa, M.~Hassan, N.~Bergmann, and W.~Hu,
  ``Keh-gait: Using kinetic energy harvesting for gait-based user
  authentication systems,'' \emph{IEEE Transactions on Mobile Computing},
  vol.~18, no.~1, pp. 139--152, 2018.

\bibitem{kolokas2019gait}
N.~Kolokas, S.~Krinidis, A.~Drosou, D.~Ioannidis, and D.~Tzovaras, ``Gait
  matching by mapping wearable to camera privacy-preserving recordings:
  Experimental comparison of multiple settings,'' in \emph{2019 6th
  International Conference on Control, Decision and Information Technologies
  (CoDIT)}.\hskip 1em plus 0.5em minus 0.4em\relax IEEE, 2019, pp. 338--343.

\bibitem{ferreira2017user}
A.~Ferreira, G.~Santos, A.~Rocha, and S.~Goldenstein, ``User-centric
  coordinates for applications leveraging 3-axis accelerometer data,''
  \emph{IEEE Sensors Journal}, vol.~17, no.~16, pp. 5231--5243, 2017.

\bibitem{sun2018artificial}
Y.~Sun and B.~Lo, ``An artificial neural network framework for gait-based
  biometrics,'' \emph{IEEE journal of biomedical and health informatics},
  vol.~23, no.~3, pp. 987--998, 2018.

\bibitem{ZhuQXZS19}
T.~Zhu, Z.~Qu, H.~Xu, J.~Zhang, Z.~Shao, Y.~Chen, S.~Prabhakar, and J.~Yang,
  ``Riskcog: Unobtrusive real-time user authentication on mobile devices in the
  wild,'' \emph{IEEE Transactions on Mobile Computing}, 2019.

\bibitem{monrose2000keystroke}
F.~Monrose and A.~D. Rubin, ``Keystroke dynamics as a biometric for
  authentication,'' \emph{Future Generation computer systems}, vol.~16, no.~4,
  pp. 351--359, 2000.

\bibitem{joyce1990identity}
R.~Joyce and G.~Gupta, ``Identity authentication based on keystroke
  latencies,'' \emph{Communications of the ACM}, vol.~33, no.~2, pp. 168--176,
  1990.

\bibitem{mcloughlin2009keypress}
I.~V. McLoughlin \emph{et~al.}, ``Keypress biometrics for user validation in
  mobile consumer devices,'' in \emph{2009 IEEE 13th International Symposium on
  Consumer Electronics}.\hskip 1em plus 0.5em minus 0.4em\relax IEEE, 2009, pp.
  280--284.

\bibitem{zahid2009keystroke}
S.~Zahid, M.~Shahzad, S.~A. Khayam, and M.~Farooq, ``Keystroke-based user
  identification on smart phones,'' in \emph{International Workshop on Recent
  advances in intrusion detection}.\hskip 1em plus 0.5em minus 0.4em\relax
  Springer, 2009, pp. 224--243.

\bibitem{urtiga2011keystroke}
E.~V.~C. Urtiga and E.~D. Moreno, ``Keystroke-based biometric authentication in
  mobile devices,'' \emph{IEEE Latin America Transactions}, vol.~9, no.~3, pp.
  368--375, 2011.

\bibitem{hwang2009keystroke}
S.-s. Hwang, S.~Cho, and S.~Park, ``Keystroke dynamics-based authentication for
  mobile devices,'' \emph{Computers \& Security}, vol.~28, no. 1-2, pp. 85--93,
  2009.

\bibitem{wu2015smartphone}
J.-S. Wu, W.-C. Lin, C.-T. Lin, and T.-E. Wei, ``Smartphone continuous
  authentication based on keystroke and gesture profiling,'' in \emph{2015
  International Carnahan Conference on Security Technology (ICCST)}.\hskip 1em
  plus 0.5em minus 0.4em\relax IEEE, 2015, pp. 191--197.

\bibitem{giuffrida2014sensed}
C.~Giuffrida, K.~Majdanik, M.~Conti, and H.~Bos, ``I sensed it was you:
  authenticating mobile users with sensor-enhanced keystroke dynamics,'' in
  \emph{International Conference on Detection of Intrusions and Malware, and
  Vulnerability Assessment}.\hskip 1em plus 0.5em minus 0.4em\relax Springer,
  2014, pp. 92--111.

\bibitem{inguanez2016securing}
F.~Inguanez and S.~Ahmadi, ``Securing smartphones via typing heat maps,'' in
  \emph{2016 IEEE 6th International Conference on Consumer Electronics-Berlin
  (ICCE-Berlin)}.\hskip 1em plus 0.5em minus 0.4em\relax IEEE, 2016, pp.
  193--197.

\bibitem{anusas2019strengthening}
T.~Anusas-amornkul, ``Strengthening password authentication using keystroke
  dynamics and smartphone sensors,'' in \emph{Proceedings of the 9th
  International Conference on Information Communication and Management}.\hskip
  1em plus 0.5em minus 0.4em\relax ACM, 2019, pp. 70--74.

\bibitem{shankar2019intelligent}
V.~Shankar and K.~Singh, ``An intelligent scheme for continuous authentication
  of smartphone using deep auto encoder and softmax regression model easy for
  user brain,'' \emph{IEEE Access}, vol.~7, pp. 48\,645--48\,654, 2019.

\bibitem{derawi2010unobtrusive}
M.~O. Derawi, C.~Nickel, P.~Bours, and C.~Busch, ``Unobtrusive
  user-authentication on mobile phones using biometric gait recognition,'' in
  \emph{2010 Sixth International Conference on Intelligent Information Hiding
  and Multimedia Signal Processing}.\hskip 1em plus 0.5em minus 0.4em\relax
  IEEE, 2010, pp. 306--311.

\bibitem{stanciu2016effectiveness}
V.-D. Stanciu, R.~Spolaor, M.~Conti, and C.~Giuffrida, ``On the effectiveness
  of sensor-enhanced keystroke dynamics against statistical attacks,'' in
  \emph{Proceedings of the Sixth ACM Conference on Data and Application
  Security and Privacy}.\hskip 1em plus 0.5em minus 0.4em\relax ACM, 2016, pp.
  105--112.

\bibitem{sitova2015hmog}
Z.~Sitov{\'a}, J.~{\v{S}}ed{\v{e}}nka, Q.~Yang, G.~Peng, G.~Zhou, P.~Gasti, and
  K.~S. Balagani, ``Hmog: New behavioral biometric features for continuous
  authentication of smartphone users,'' \emph{IEEE Transactions on Information
  Forensics and Security}, vol.~11, no.~5, pp. 877--892, 2015.

\bibitem{BuriroCGF18}
A.~Buriro, B.~Crispo, S.~Gupta, and F.~Del~Frari, ``{DIALERAUTH:} {A}
  motion-assisted touch-based smartphone user authentication scheme,'' in
  \emph{Proceedings of the Eighth ACM Conference on Data and Application
  Security and Privacy}, 2018, pp. 267--276.

\bibitem{mondal2015swipe}
S.~Mondal and P.~Bours, ``Swipe gesture based continuous authentication for
  mobile devices,'' in \emph{2015 International Conference on Biometrics
  (ICB)}.\hskip 1em plus 0.5em minus 0.4em\relax IEEE, 2015, pp. 458--465.

\bibitem{nohara2016personal}
T.~Nohara and R.~Uda, ``Personal identification by flick input using
  self-organizing maps with acceleration sensor and gyroscope,'' in
  \emph{Proceedings of the 10th International Conference on Ubiquitous
  Information Management and Communication}.\hskip 1em plus 0.5em minus
  0.4em\relax ACM, 2016, p.~58.

\bibitem{lin2012new}
C.-C. Lin, C.-C. Chang, D.~Liang, and C.-H. Yang, ``A new non-intrusive
  authentication method based on the orientation sensor for smartphone users,''
  in \emph{2012 IEEE Sixth International Conference on Software Security and
  Reliability}.\hskip 1em plus 0.5em minus 0.4em\relax IEEE, 2012, pp.
  245--252.

\bibitem{lu2015safeguard}
L.~Lu and Y.~Liu, ``Safeguard: User reauthentication on smartphones via
  behavioral biometrics,'' \emph{IEEE Transactions on Computational Social
  Systems}, vol.~2, no.~3, pp. 53--64, 2015.

\bibitem{jain2015exploring}
A.~Jain and V.~Kanhangad, ``Exploring orientation and accelerometer sensor data
  for personal authentication in smartphones using touchscreen gestures,''
  \emph{Pattern recognition letters}, vol.~68, pp. 351--360, 2015.

\bibitem{shih2015flick}
D.-H. Shih, C.-M. Lu, and M.-H. Shih, ``A flick biometric authentication
  mechanism on mobile devices,'' in \emph{2015 International Conference on
  Informative and Cybernetics for Computational Social Systems (ICCSS)}.\hskip
  1em plus 0.5em minus 0.4em\relax IEEE, 2015, pp. 31--33.

\bibitem{saevanee2008user}
H.~Saevanee and P.~Bhatarakosol, ``User authentication using combination of
  behavioral biometrics over the touchpad acting like touch screen of mobile
  device,'' in \emph{2008 International Conference on Computer and Electrical
  Engineering}.\hskip 1em plus 0.5em minus 0.4em\relax IEEE, 2008, pp. 82--86.

\bibitem{xu2014towards}
H.~Xu, Y.~Zhou, and M.~R. Lyu, ``Towards continuous and passive authentication
  via touch biometrics: An experimental study on smartphones,'' in \emph{10th
  Symposium On Usable Privacy and Security ($\{$SOUPS$\}$ 2014)}, 2014, pp.
  187--198.

\bibitem{nixon2016slowmo}
K.~W. Nixon, X.~Chen, Z.-H. Mao, and Y.~Chen, ``Slowmo-enhancing mobile
  gesture-based authentication schemes via sampling rate optimization,'' in
  \emph{2016 21st Asia and South Pacific Design Automation Conference
  (ASP-DAC)}.\hskip 1em plus 0.5em minus 0.4em\relax IEEE, 2016, pp. 462--467.

\bibitem{nader2015designing}
J.~Nader, A.~Alsadoon, P.~Prasad, A.~Singh, and A.~Elchouemi, ``Designing
  touch-based hybrid authentication method for smartphones,'' \emph{Procedia
  Computer Science}, vol.~70, pp. 198--204, 2015.

\bibitem{antal2016biometric}
M.~Antal and L.~Z. Szab{\'o}, ``Biometric authentication based on touchscreen
  swipe patterns,'' \emph{Procedia Technology}, vol.~22, pp. 862--869, 2016.

\bibitem{primo2015music}
A.~Primo and V.~V. Phoha, ``Music and images as contexts in a context-aware
  touch-based authentication system,'' in \emph{2015 IEEE 7th International
  Conference on Biometrics Theory, Applications and Systems (BTAS)}.\hskip 1em
  plus 0.5em minus 0.4em\relax IEEE, 2015, pp. 1--7.

\bibitem{feng2014tips}
T.~Feng, J.~Yang, Z.~Yan, E.~M. Tapia, and W.~Shi, ``Tips: Context-aware
  implicit user identification using touch screen in uncontrolled
  environments,'' in \emph{Proceedings of the 15th Workshop on Mobile Computing
  Systems and Applications}.\hskip 1em plus 0.5em minus 0.4em\relax ACM, 2014,
  p.~9.

\bibitem{shen2015performance}
C.~Shen, Y.~Zhang, X.~Guan, and R.~A. Maxion, ``Performance analysis of
  touch-interaction behavior for active smartphone authentication,'' \emph{IEEE
  Transactions on Information Forensics and Security}, vol.~11, no.~3, pp.
  498--513, 2015.

\bibitem{syed2019touch}
Z.~Syed, J.~Helmick, S.~Banerjee, and B.~Cukic, ``Touch gesture-based
  authentication on mobile devices: The effects of user posture, device size,
  configuration, and inter-session variability,'' \emph{Journal of Systems and
  Software}, vol. 149, pp. 158--173, 2019.

\bibitem{rauen2018gesture}
Z.~I. Rauen, F.~Anjomshoa, and B.~Kantarci, ``Gesture and sociability-based
  continuous authentication on smart mobile devices,'' in \emph{Proceedings of
  the 16th ACM International Symposium on Mobility Management and Wireless
  Access}.\hskip 1em plus 0.5em minus 0.4em\relax ACM, 2018, pp. 51--58.

\bibitem{rocha2019continuous}
R.~Rocha, D.~Carneiro, R.~Costa, and C.~Analide, ``Continuous authentication in
  mobile devices using behavioral biometrics,'' in \emph{International
  Symposium on Ambient Intelligence}.\hskip 1em plus 0.5em minus 0.4em\relax
  Springer, 2019, pp. 191--198.

\bibitem{mondal2015continuous}
S.~Mondal and P.~Bours, ``Continuous authentication and identification for
  mobile devices: Combining security and forensics,'' in \emph{2015 IEEE
  International Workshop on Information Forensics and Security (WIFS)}.\hskip
  1em plus 0.5em minus 0.4em\relax IEEE, 2015, pp. 1--6.

\bibitem{alsulaiman2008user}
F.~A. Alsulaiman, J.~Cha, and A.~El~Saddik, ``User identification based on
  handwritten signatures with haptic information,'' in \emph{International
  Conference on Human Haptic Sensing and Touch Enabled Computer
  Applications}.\hskip 1em plus 0.5em minus 0.4em\relax Springer, 2008, pp.
  114--121.

\bibitem{cai2011touchlogger}
L.~Cai and H.~Chen, ``Touchlogger: Inferring keystrokes on touch screen from
  smartphone motion.'' \emph{HotSec}, vol.~11, no. 2011, p.~9, 2011.

\bibitem{miguel2016predicting}
O.~Miguel-Hurtado, S.~V. Stevenage, C.~Bevan, and R.~Guest, ``Predicting sex as
  a soft-biometrics from device interaction swipe gestures,'' \emph{Pattern
  Recognition Letters}, vol.~79, pp. 44--51, 2016.

\bibitem{bevan2016different}
C.~Bevan and D.~S. Fraser, ``Different strokes for different folks? revealing
  the physical characteristics of smartphone users from their swipe gestures,''
  \emph{International Journal of Human-Computer Studies}, vol.~88, pp. 51--61,
  2016.

\bibitem{azenkot2012passchords}
S.~Azenkot, K.~Rector, R.~Ladner, and J.~Wobbrock, ``Passchords: secure
  multi-touch authentication for blind people,'' in \emph{Proceedings of the
  14th international ACM SIGACCESS conference on Computers and
  accessibility}.\hskip 1em plus 0.5em minus 0.4em\relax ACM, 2012, pp.
  159--166.

\bibitem{khan2014itus}
H.~Khan, A.~Atwater, and U.~Hengartner, ``Itus: an implicit authentication
  framework for android,'' in \emph{Proceedings of the 20th annual
  international conference on Mobile computing and networking}.\hskip 1em plus
  0.5em minus 0.4em\relax ACM, 2014, pp. 507--518.

\bibitem{miguel2016interaction}
O.~Miguel-Hurtado, R.~Blanco-Gonzalo, R.~Guest, and C.~Lunerti, ``Interaction
  evaluation of a mobile voice authentication system,'' in \emph{2016 IEEE
  International Carnahan Conference on Security Technology (ICCST)}.\hskip 1em
  plus 0.5em minus 0.4em\relax IEEE, 2016, pp. 1--8.

\bibitem{zhang2017hearing}
L.~Zhang, S.~Tan, and J.~Yang, ``Hearing your voice is not enough: An
  articulatory gesture based liveness detection for voice authentication,'' in
  \emph{Proceedings of the 2017 ACM SIGSAC Conference on Computer and
  Communications Security}.\hskip 1em plus 0.5em minus 0.4em\relax ACM, 2017,
  pp. 57--71.

\bibitem{lu2019lip}
L.~Lu, J.~Yu, Y.~Chen, H.~Liu, Y.~Zhu, L.~Kong, and M.~Li, ``Lip reading-based
  user authentication through acoustic sensing on smartphones,'' \emph{IEEE/ACM
  Transactions on Networking}, vol.~27, no.~1, pp. 447--460, 2019.

\bibitem{wang2019voicepop}
Q.~Wang, X.~Lin, M.~Zhou, Y.~Chen, C.~Wang, Q.~Li, and X.~Luo, ``Voicepop: A
  pop noise based anti-spoofing system for voice authentication on
  smartphones,'' in \emph{IEEE INFOCOM 2019-IEEE Conference on Computer
  Communications}.\hskip 1em plus 0.5em minus 0.4em\relax IEEE, 2019, pp.
  2062--2070.

\bibitem{yan2016usable}
Z.~Yan and S.~Zhao, ``A usable authentication system based on personal voice
  challenge,'' in \emph{2016 International Conference on Advanced Cloud and Big
  Data (CBD)}.\hskip 1em plus 0.5em minus 0.4em\relax IEEE, 2016, pp. 194--199.

\bibitem{kim2008multimodal}
D.-S. Kim and K.-S. Hong, ``Multimodal biometric authentication using teeth
  image and voice in mobile environment,'' \emph{IEEE Transactions on Consumer
  Electronics}, vol.~54, no.~4, pp. 1790--1797, 2008.

\bibitem{johnson2013secure}
R.~Johnson, W.~J. Scheirer, and T.~E. Boult, ``Secure voice-based
  authentication for mobile devices: vaulted voice verification,'' in
  \emph{Biometric and Surveillance Technology for Human and Activity
  Identification X}, vol. 8712.\hskip 1em plus 0.5em minus 0.4em\relax
  International Society for Optics and Photonics, 2013, p. 87120P.

\bibitem{zhang2016voicelive}
L.~Zhang, S.~Tan, J.~Yang, and Y.~Chen, ``Voicelive: A phoneme localization
  based liveness detection for voice authentication on smartphones,'' in
  \emph{Proceedings of the 2016 ACM SIGSAC Conference on Computer and
  Communications Security}.\hskip 1em plus 0.5em minus 0.4em\relax ACM, 2016,
  pp. 1080--1091.

\bibitem{atal1974effectiveness}
B.~S. Atal, ``Effectiveness of linear prediction characteristics of the speech
  wave for automatic speaker identification and verification,'' \emph{the
  Journal of the Acoustical Society of America}, vol.~55, no.~6, pp.
  1304--1312, 1974.

\bibitem{reynolds2002overview}
D.~A. Reynolds, ``An overview of automatic speaker recognition technology,'' in
  \emph{2002 IEEE International Conference on Acoustics, Speech, and Signal
  Processing}, vol.~4.\hskip 1em plus 0.5em minus 0.4em\relax IEEE, 2002, pp.
  IV--4072.

\bibitem{kinnunen2010overview}
T.~Kinnunen and H.~Li, ``An overview of text-independent speaker recognition:
  From features to supervectors,'' \emph{Speech communication}, vol.~52, no.~1,
  pp. 12--40, 2010.

\bibitem{gofman2016multimodal}
M.~I. Gofman, S.~Mitra, T.-H.~K. Cheng, and N.~T. Smith, ``Multimodal
  biometrics for enhanced mobile device security,'' \emph{Communications of the
  ACM}, vol.~59, no.~4, pp. 58--65, 2016.

\bibitem{khamis2016gazetouchpass}
M.~Khamis, F.~Alt, M.~Hassib, E.~von Zezschwitz, R.~Hasholzner, and A.~Bulling,
  ``Gazetouchpass: Multimodal authentication using gaze and touch on mobile
  devices,'' in \emph{Proceedings of the 2016 CHI Conference Extended Abstracts
  on Human Factors in Computing Systems}.\hskip 1em plus 0.5em minus
  0.4em\relax ACM, 2016, pp. 2156--2164.

\bibitem{saevanee2015continuous}
H.~Saevanee, N.~Clarke, S.~Furnell, and V.~Biscione, ``Continuous user
  authentication using multi-modal biometrics,'' \emph{Computers \& Security},
  vol.~53, pp. 234--246, 2015.

\bibitem{neal2015mobile}
T.~J. Neal, D.~L. Woodard, and A.~D. Striegel, ``Mobile device application,
  bluetooth, and wi-fi usage data as behavioral biometric traits,'' in
  \emph{2015 IEEE 7th International Conference on Biometrics Theory,
  Applications and Systems (BTAS)}.\hskip 1em plus 0.5em minus 0.4em\relax
  IEEE, 2015, pp. 1--6.

\bibitem{raja2015multi}
K.~B. Raja, R.~Raghavendra, M.~Stokkenes, and C.~Busch, ``Multi-modal
  authentication system for smartphones using face, iris and periocular,'' in
  \emph{2015 International Conference on Biometrics (ICB)}.\hskip 1em plus
  0.5em minus 0.4em\relax IEEE, 2015, pp. 143--150.

\bibitem{stokkenes2017feature}
M.~Stokkenes, R.~Ramachandra, K.~B. Raja, M.~K. Sigaard, and C.~Busch,
  ``Feature level fused templates for multi-biometric system on smartphones,''
  in \emph{2017 5th International Workshop on Biometrics and Forensics
  (IWBF)}.\hskip 1em plus 0.5em minus 0.4em\relax IEEE, 2017, pp. 1--5.

\bibitem{tiong2017multimodal}
L.~C.~O. Tiong, S.~T. Kim, and Y.~M. Ro, ``Multimodal face biometrics by using
  convolutional neural networks,'' \emph{Journal of Korea Multimedia Society},
  vol.~20, no.~2, pp. 170--178, 2017.

\bibitem{lamiche2019continuous}
I.~Lamiche, G.~Bin, Y.~Jing, Z.~Yu, and A.~Hadid, ``A continuous smartphone
  authentication method based on gait patterns and keystroke dynamics,''
  \emph{Journal of Ambient Intelligence and Humanized Computing}, vol.~10,
  no.~11, pp. 4417--4430, 2019.

\bibitem{pang2019mineauth}
X.~Pang, L.~Yang, M.~Liu, and J.~Ma, ``Mineauth: Mining behavioural habits for
  continuous authentication on a smartphone,'' in \emph{Australasian Conference
  on Information Security and Privacy}.\hskip 1em plus 0.5em minus 0.4em\relax
  Springer, 2019, pp. 533--551.

\bibitem{acien2019multilock}
A.~Acien, A.~Morales, R.~Vera-Rodriguez, J.~Fierrez, and R.~Tolosana,
  ``Multilock: Mobile active authentication based on multiple biometric and
  behavioral patterns,'' in \emph{1st International Workshop on Multimodal
  Understanding and Learning for Embodied Applications}.\hskip 1em plus 0.5em
  minus 0.4em\relax ACM, 2019, pp. 53--59.

\bibitem{KayacikJBAM14}
H.~G. Kayacik, M.~Just, L.~Baillie, D.~Aspinall, and N.~Micallef, ``Data driven
  authentication: On the effectiveness of user behaviour modelling with mobile
  device sensors,'' \emph{CoRR}, vol. abs/1410.7743, 2014.

\bibitem{ZhuWWZ13}
J.~Zhu, P.~Wu, X.~Wang, and J.~Zhang, ``Sensec: Mobile security through passive
  sensing,'' in \emph{International Conference on Computing, Networking and
  Communications, {ICNC}}, 2013, pp. 1128--1133.

\bibitem{FenuM18}
G.~Fenu and M.~Marras, ``Controlling user access to cloud-connected mobile
  applications by means of biometrics,'' \emph{{IEEE} Cloud Computing}, vol.~5,
  no.~4, pp. 47--57, 2018.

\bibitem{lin2018tdsd}
H.~Lin, J.~Liu, and Q.~Li, ``Tdsd: A touch dynamic and sensor data based
  approach for continuous user authentication.'' in \emph{PACIS}, 2018, p. 294.

\bibitem{volaka2019towards}
H.~C. Volaka, G.~Alptekin, O.~E. Basar, M.~Isbilen, and O.~D. Incel, ``Towards
  continuous authentication on mobile phones using deep learning models,''
  \emph{Procedia Computer Science}, vol. 155, pp. 177--184, 2019.

\bibitem{ShenYYLG16}
C.~Shen, T.~Yu, S.~Yuan, Y.~Li, and X.~Guan, ``Performance analysis of
  motion-sensor behavior for user authentication on smartphones,''
  \emph{Sensors}, vol.~16, no.~3, p. 345, 2016.

\bibitem{centeno2018mobile}
M.~P. Centeno, Y.~Guan, and A.~van Moorsel, ``Mobile based continuous
  authentication using deep features,'' in \emph{Proceedings of the 2nd
  International Workshop on Embedded and Mobile Deep Learning}.\hskip 1em plus
  0.5em minus 0.4em\relax ACM, 2018, pp. 19--24.

\bibitem{MoseniaSRJ17}
A.~Mosenia, S.~Sur{-}Kolay, A.~Raghunathan, and N.~K. Jha, ``{CABA:} continuous
  authentication based on bioaura,'' \emph{{IEEE} Trans. Computers}, vol.~66,
  no.~5, pp. 759--772, 2017.

\bibitem{BaQFR19}
Z.~Ba, Z.~Qin, X.~Fu, and K.~Ren, ``{CIM:} camera in motion for smartphone
  authentication,'' \emph{{IEEE} Trans. Information Forensics and Security},
  vol.~14, no.~11, pp. 2987--3002, 2019.

\bibitem{jain2006biometrics}
A.~K. Jain, A.~Ross, and S.~Pankanti, ``Biometrics: a tool for information
  security,'' \emph{IEEE transactions on information forensics and security},
  vol.~1, no.~2, pp. 125--143, 2006.

\bibitem{rahman2014seeing}
F.~Rahman, M.~O. Gani, G.~M.~T. Ahsan, and S.~I. Ahamed, ``Seeing beyond
  visibility: A four way fusion of user authentication for efficient usable
  security on mobile devices,'' in \emph{2014 IEEE Eighth International
  Conference on Software Security and Reliability-Companion}.\hskip 1em plus
  0.5em minus 0.4em\relax IEEE, 2014, pp. 121--129.

\bibitem{LiB18}
G.~Li and P.~Bours, ``Studying wifi and accelerometer data based authentication
  method on mobile phones,'' in \emph{Proceedings of the 2nd International
  Conference on Biometric Engineering and Applications}, 2018, pp. 18--23.

\bibitem{raja2015fusion}
K.~B. Raja, R.~Raghavendra, M.~Stokkenes, and C.~Busch, ``Fusion of face and
  periocular information for improved authentication on smartphones,'' in
  \emph{2015 18th International Conference on Information Fusion
  (Fusion)}.\hskip 1em plus 0.5em minus 0.4em\relax IEEE, 2015, pp. 2115--2120.

\end{thebibliography}

\end{document}